\documentclass[aps,amsmath,amssymb,amsfonts,twocolumn,superscriptaddress]{revtex4-2}
\usepackage{bm}
\usepackage{color,graphicx}
\usepackage[version=4]{mhchem}
\usepackage{mathrsfs}

\usepackage{xcolor,hyperref}
\hypersetup{
   colorlinks,
   linkcolor={red!80!black},
   citecolor={blue!50!black},
   urlcolor={blue!80!black}
}
\usepackage{cleveref}

\newcommand{\p}{\partial}
\newcommand{\diff}{\mathrm{d}}
\newcommand{\ttz}{\rightarrow 0}
\newcommand{\tti}{\rightarrow \infty}

\crefname{equation}{Eq.}{Eqs.}
\crefname{figure}{Fig.}{Figs.}
\crefname{section}{Sec.}{Secs.}
\crefrangeformat{equation}{Eqs.~#3(#1)#4--#5(#2)#6}

\begin{document}

\title{Thermoelectric effect and temperature-gradient-driven electrokinetic flow of electrolyte solutions in charged nanocapillaries}
\thanks{Cite this: W. Zhang, et al. Int. J. Heat Mass Transfer 143 (2019) 118569.}
	
\author{Wenyao Zhang}
\author{Qiuwang Wang}
\author{Min Zeng}
\author{Cunlu Zhao}
\email{mclzhao@xjtu.edu.cn}
\affiliation{MOE Key Laboratory of Thermo-Fluid Science and Engineering, School of Energy and Power Engineering, Xi'an Jiaotong University, Xi'an 710049, China}
	
\begin{abstract}
A systematic theoretical study of thermoelectric effect and temperature-gradient-driven electrokinetic flow of electrolyte solutions in charged nanocapillaries is presented. The study is based on a semianalytical model developed by simultaneously solving the non-isothermal Poisson-Nernst-Planck-Navier-Stokes equations with the lubrication theory. Particularly, this paper clarifies the interplay and relative importance of the thermoelectric mechanisms due to (a) the convective transport of ions caused by the fluid flow, (b) the dependence of ion electrophoretic mobility on temperature, (c) the difference in the intrinsic Soret coefficients of cation and anion. Additionally, synergy conditions for the three thermoelectric mechanisms to fully cooperate are proposed for thermo-phobic/philic electrolytes. The temperature-gradient-driven electrokinetic flow is shown to be a nearly unidirectional flow whose axial velocity profiles vary with the axial location. Also, the flow can be regarded as a consequence of the counteraction or cooperation between a thermoelectric-field-driven electroosmotic flow and a thermo-osmotic flow driven by the osmotic pressure gradient and dielectric body force. Moreover, the Seebeck coefficient and the fluid average velocity are demonstrated to be affected by electrolyte-related parameters. The results are beneficial for under- standing the temperature-gradient-driven electrokinetic transport in nanocapillaries and also serve as theoretical foundation for the design of low-grade waste heat recovery devices and thermoosmotic pumps.
\end{abstract}

\maketitle

\section{Introduction}\label{sec:intro}

Electrokinetic transport in nanoscale channels and pores have a variety of applications in the realms of science and engineering, such as electrokinetic energy conservation \cite{Heyden2006}, osmotic energy conversion \cite{Siria2013}, water desalination \cite{Picallo2013}, particle separation \cite{Inglis2011}, and nanofluidic transistors \cite{Karnik2005} etc. These applications can be achieved by applying one or more types of thermodynamic forces (i.e., gradients in pressure, concentration, electric potential or temperature). Specifically, the application of an axial temperature gradient along the charged nanocapillaries can produce an thermoelectric potential as well as an accompanying electrokinetic flow \cite{derjaguin1987,dietzel2017}. As a category of non-isothermal electrokinetic transport process, this phenomenon is believed to have diverse promising technological applications, such as low-grade waste heat recovery, liquid pumping, water treatment, etc \cite{dietzel2016,dietzel2017,barragan2017}.

Over the past two decades, most studies focus on the electrokinetic transport through micro-/nano-channels under isothermal conditions, such as electroosmosis \cite{zhao2008,Qi2018} and streaming potential \cite{Heyden2006}, etc. These phenomena have been widely investigated in the literature; their theories have been well developed and corresponding physical mechanisms have been well understood. Yet, when it comes to the non-isothermal electrokinetic transport in nanochannels, the relevant studies are rare, and generally there is a lack of a clear understanding of its physics. This status is mainly due to the fact that the theory development of the non-isothermal electrokinetic transport in nanochannels is complicated by at least three aspects of non-isothermal effects \cite{dietzel2017}: (\romannumeral1) The energy equation must be solved simultaneously with other electrokinetic governing equations to obtain the temperature distribution \cite{maynes2004,Xuan2004a,Sadeghi2010}, which is the basis for evaluating local physical properties and analyzing the thermophoretic ion motion (also termed intrinsic Soret effect); (\romannumeral2) Physical properties of electrolyte solutions (e.g., viscosity, permittivity, ion electrophoretic mobility and ion diffusivity) are temperature-dependent in nature, and this may give rise to various effects, such as a dielectric body force due to the temperature dependence of the permittivity \cite{derjaguin1987}; (\romannumeral3) The thermophoretic ion fluxes should be included into the Nernst-Planck equation other than the conventional diffusive, convective and electromigrative fluxes \cite{degroot1984}. These non-isothermal effects make it formidably difficult to derive an exact or even approximate analytical model for the non-isothermal electrokinetic transport in nanochannels due to the involvement of temperature coupling and extra nonlinearity.

Currently, studies on the non-isothermal electrokinetic transport in nanochannels are relatively scarce. Most of the relevant studies focus on the heat transfer and effects of the Joule heating or viscous dissipation on electroosmotic flow (EOF) in microchannels \cite{maynes2004,Xuan2004a,Sadeghi2010}. Additionally, Researchers also pay attention to effects of temperature gradient on the classic electrokinetic transport in micro/nanochannels. For example, the temperature gradient was found to have a substantial effect on the flow rate of pressure-driven electrokinetic flow in a microchannel \cite{ghonge2013}, the ion selectivity of nanochannels \cite{wood2016,benneker2017} and the performance of nanofluidic reverse electro-dialysis system \cite{Long2018,Long2019}. In nanochannels or nanopores, there exists another category of non-isothermal electrokinetic transport process driven by the temperature gradient alone (being referred to as the temperature-gradient-driven electrokinetic transport in this paper), which is the focus of this paper.

The study of the temperature-gradient-driven electrokinetic transport in micro/nanopores can be traced back to the middle of last century. The early work by Derjaguin and Sidorenkov \cite{derjaguin1941} studied the temperature-gradient-driven flow in a porous glass and reported an expression for the velocity in terms of the excess enthalpy density, and this excess enthalpy density was further found to be related to the electric double layer (EDL). Kobatake and Fujita \cite{kobatake1964} developed a model to analyze the temperature-gradient-driven electrokinetic flow in charged membrane pores to interpret the relevant experimental data \cite{carr1962}. Their model neglected the contribution of osmotic pressure and adopted the assumptions of (\romannumeral1) the Debye-H\"uckel linearization and (\romannumeral2) thin EDL. During 1970--90s, Tasaka~\textit{et al.} performed comprehensive investigations of the temperature-gradient-driven electrokinetic transport of electrolyte solutions across ion-exchange membranes \cite{tasaka1978,tasaka1986,Tasaka1992a}. In their studies, the temperature-gradient-driven fluxes of electrolyte solutions across the ion-exchange membrane were measured under various conditions and further used to infer the cross coefficients in the Onsager reciprocal relations and the transport numbers. However, these studies paid no attention to the transport mechanisms within the membranes (which consist of dense cross-linked polymers and thus free volume forms inside; such free volume can be equivalently treated as effective ``pores'' whose radius distribution is generally determined by low-temperature nitrogen adsorption experiment), and thus did not relate the cross coefficients and transport number to the membrane properties (such as charge density and effective pore radius) and electrolyte properties (e.g., ion diffusivity and ion intrinsic Soret coefficient). Quite recently, due to the promising applications in waste heat recovery \cite{Sandbakk2013,Jokinen2016,Xie2018} and fluid delivery \cite{Maheedhara2018}, temperature-gradient-driven electrokinetic transport of electrolyte solutions in nanochannels or nanopores receive the renewed interests. The studies now focus on more detailed characteristics of the temperature-gradient-driven electrokinetic transport. Bregulla~\textit{et al.}~\cite{bregulla2016} reported the first experimental observation of the temperature-gradient-driven electrokinetic flow field with the micro-particle image velocimetry. The details of flow field in a thin electrolyte solution film confined between two charged plates was obtained. Immediately, Dietzel and Hardt~\cite{dietzel2017} presented a detailed theoretical analysis of the temperature-gradient-driven electrokinetic transport in slit mirco/nanochannels. In their work, the effects of $ \zeta $ potential and channel size on the thermoosmotic flow (TOF) and the corresponding streaming electric field were discussed in details.

Although relevant studies have been carried out in a span of more than seventy years for the temperature-gradient-driven electrokinetic transport in channels, the effort to acquire a detailed understanding of this phenomenon only emerges in recent studies \cite{dietzel2016,dietzel2017}. These studies  performed a study of the thermoelectric potential and the flow of electrolyte solutions in slit channels with a hydrodynamic approach. In addition to the well-known thermoelectric effect resulting from the difference in the intrinsic Soret coefficients of cations and anions, they identified two extra thermoelectric effects in confined electrolyte solutions, i.e., thermoelectric potential due to (1) the liquid flow, and (2) the temperature dependence of the ion electrophoretic mobility in conjunction with ultra-confined channels. However, their model for the temperature-gradient-driven electrokinetic transport of electrolyte solutions in these studies is suitable for two-dimensional slit channels. Actually, a capillary model for the temperature-gradient-driven electrokinetic transport of electrolyte solutions is more relevant since a single capillary is the basis for the understanding and prediction of transport characteristics in the practically used porous media and (ion exchange or nanofluidic) membranes. Unfortunately, the current literature lacks such a capillary model. Moreover, influence of various model parameters on the temperature-gradient-driven electrokinetic transport (e.g., the flow behavior and thermoelectric potential) in charged nanocapillaries are yet to be investigated.

This work aims at developing a model for the temperature-gradient-driven electrokinetic transport in charged nanocapillaries. For this purpose, we directly solve the nonisothermal Poisson-Nernst-Planck-Navier-Stokes equations via the lubrication approximation and derive semi-analytical expressions for the flow field, average fluid velocity and thermoelectric potential in charged nanocapillaries. Notably, the derived semi-analytical solutions are in agreement with the classic isothermal capillary pore model \cite{Gross1968,peters2016} when the temperature gradient vanishes, and also can exactly recover widely-accepted theories in limiting cases. Basing on our semi-analytical solutions, we discuss the interplay among three thermoelectric effects in depth, and clarify the relative importance of each thermoelectric effect, and finally propose synergistic conditions for different thermoelectric effects to fully cooperate. Additionally, characteristics of the temperature-gradient-driven electrokinetic flow are analyzed in details for varying dimensionless Debye parameters (being defined later), $ \zeta $ potentials and electrolytes. To the best of our knowledge, all of these have yet to be unambiguously studied.

The rest of this paper is organized as follows. In \cref{sec:theory}, the semi-analytical model for the temperature-gradient-driven electrokinetic transport of electrolyte solutions in nanocapillaries is developed. In \cref{results}, the interplay among the different physical mechanisms of thermoelectric effect and fluid flow is discussed, meanwhile, the effects of various model parameters on the Seebeck coefficient and the fluid flow are elaborated. Finally, our findings are summarized in \cref{sec:conclusions}.

\section{Theory}\label{sec:theory}
\begin{figure*}[!htb]
\centering
\includegraphics[width=0.75\linewidth]{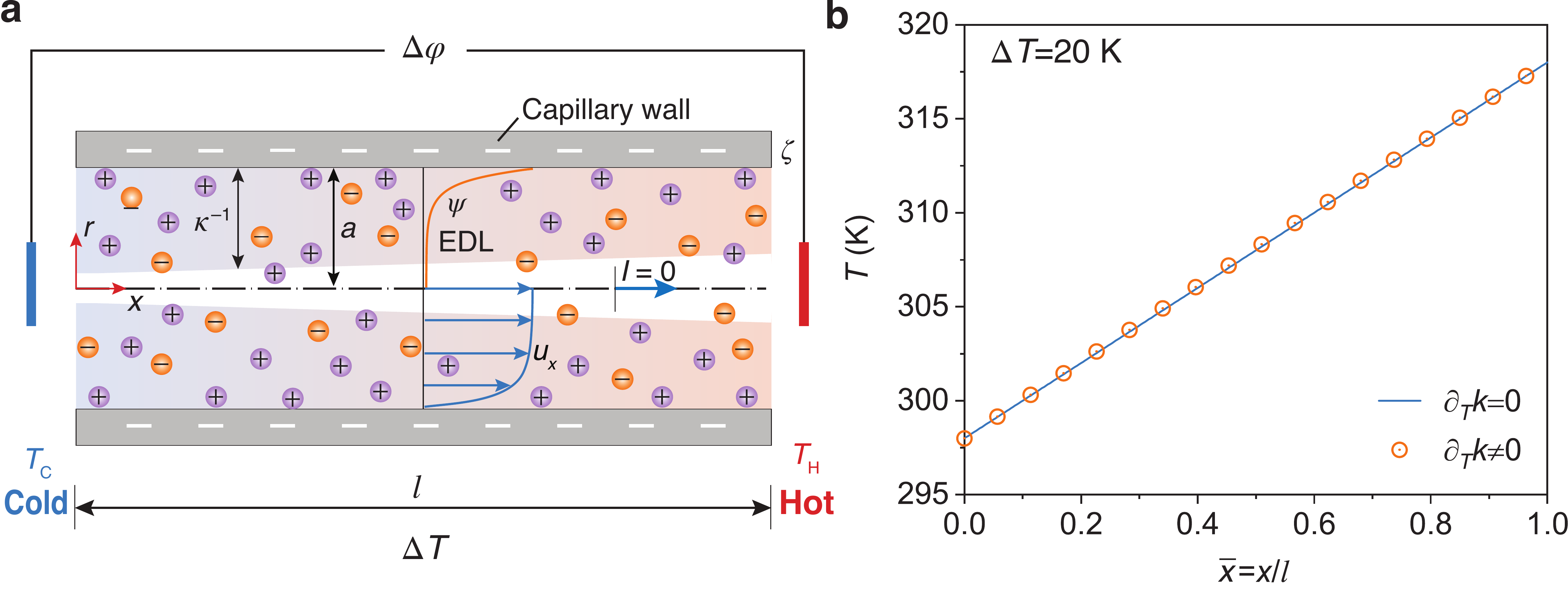}
\caption{(a) Schematic of a (nano)capillary of radius $a$ and length $l$, and both ends are submerged in reservoirs containing an aqueous solution of electrolyte. The capillary wall in contact with aqueous electrolyte acquires charges, which are screened by the ions in the EDL. The EDL is characterized by the $ \zeta $ potential on the shear plane. A temperature difference $\Delta T = T_{\mathrm{H}} - T_{\mathrm{C}} $ is applied between two ends of the capillary to induce an electric potential difference $\Delta \varphi$ and an accompanying fluid flow (whose velocity profile is schematically displayed). Due to the temperature difference, both the EDL potential, $ \psi $, and the EDL characteristic thickness ($\simeq\kappa^{-1}$) vary along axial direction $x$. The cross-sectional average electric current density, $I$, vanishes at the steady state. (b) Temperature distribution along the capillary. The linear analytical solution based on constant thermal conductivity (curve) is compared to the numerical solutions considering the temperature dependence of thermal conductivity (symbol). Here, the temperature dependent thermal conductivity of water is given in Appendix~\ref{appB}.}
\label{fig:sketch}
\end{figure*}

Fig.~\ref{fig:sketch}a schematically depicts the investigated system, which consists of a capillary with the radius of $a$ and represents an element in ion-exchange/nanofluidic membranes or porous media. The capillary filled with an aqueous electrolyte solution is connected with one large reservoir (not shown) at each capillary end. The solution in the left reservoir is assumed to be at the concentration $ n_{\infty} $ and the ambient temperature $ T_{\infty} $, while the solution in the right reservoir is assumed to be at temperature $ T_{\infty}+ \Delta T $. Therefore, there exists a temperature difference, $ \Delta T $, over the capillary length of $ l $. The capillary wall is electrically charged and an EDL in the liquid phase near the charged wall forms accordingly. Under the effect of the temperature difference, separation of charge carriers (cation and anion) occurs due to various mechanisms and thus a stable thermoelectric field is set up when the electric current vanishes (i.e., at steady state). Simultaneously, a temperature-gradient-driven electrokinetic flow is also produced by driving forces of various origins. In contrast to the conventional isothermal electrokinetic phenomena, the absolute temperature, $ T $, in the system depicted in \cref{fig:sketch}a is no longer uniform.  At a steady state, $ T $ is governed by energy equation, which at small thermal P{\'e}clet numbers and negligible viscous dissipation (see Appendix~\ref{appA}) can be written as $ \nabla\cdot(k\nabla T) =0$ with $ k $ being the thermal conductivity of the solution. Under conditions of $ a/l \ll 1 $, with the help of order-of-magnitude analysis and zero heat flux on the wall one finds that $ T $ is independent of $ r $-coordinate, resulting in $ \partial_x (k \partial_x T)=0 $. Neglecting the temperature dependence of the thermal conductivity, one obtains $ T=T_{\infty}+(x/l)\Delta T $, where $ x $ is the axial coordinate. Actually, the deviation of temperature distribution along the nanocapillary from linearity is negligible even if the temperature dependence of the thermal conductivity is taken into account (\cref{fig:sketch}b).

For a 1:1 electrolyte, at the steady state the ion transport in dilute solutions is described by the extended Nernst-Planck equation
\begin{equation}\label{nernst_planck_equation}
\nabla\cdot \bm{j}_{\pm} =0
\end{equation}
with
\begin{equation}\label{eq:density_flux}
\bm{j}_{\pm}=n_{\pm}\bm{u}-n_{\pm} D_{\pm} \left( \nabla\ln n_{\pm} \pm \frac{e }{k_{\mathrm{B}} T} \nabla \phi +  S_{\mathrm{T},\pm}\nabla T\right) 
\end{equation}
being the ion flux, where $ e $ is the elementary charge, $ k_{\mathrm{B}} $ is the Boltzmann constant, $ \bm{u}=(u_x,u_r) $ is the fluid velocity and $ \phi $ is the total electric potential. Additionally, $n_{\pm}$ and $D_{\pm}$ are the number concentration and the diffusivity of cation and anion, respectively. Moreover, $ S_{\mathrm{T},\pm} \simeq Q^{\ast}_{\pm}/(k_{\mathrm{B}} T^2) $ is the intrinsic Soret coefficient of cations and anions \cite{vigolo2010} with $ Q^{\ast}_{\pm}  $ being the ion heat of transport \cite{agar1989}.

The total electric potential $\phi$ is governed by the Poisson equation
\begin{equation}\label{poisson_equation}
\nabla
\cdot(\epsilon\nabla\phi)=-\rho_e,
\end{equation}
where $\rho_e=e(n_+-n_-)$ is the free charge density and $ \epsilon $ is the dielectric permittivity of solution.

For a steady-state incompressible flow, the fluid velocity is governed by the continuity equation, $ \nabla\cdot\bm{u}=0 $, and the extended Navier-Stokes equation
\begin{equation}\label{eq:nse}
0=-\nabla p+\nabla\cdot\mu\left[\nabla\bm{u}+ (\nabla\bm{u})^{\top}\right]-\rho_e\nabla\phi-\frac{1}{2}(\nabla\phi)^2\nabla\epsilon,
\end{equation}
where $ p $ and $ \mu $ are the pressure and the viscosity of the fluid, respectively. On the right-hand side of \cref{eq:nse}, the third term is the electrostatic body force, and the last term is the dielectric body force due to the temperature dependence of the relative permittivity \cite{derjaguin1987}. Moreover, the inertial term is neglected because of the low Reynolds numbers.

Next, in analogous to the approach used in Ref.~\cite{Alizadeh2017} we simplify the non-isothermal version of the Poisson-Nernst-Planck-Navier-Stokes (PNPNS) equations given by \crefrange{nernst_planck_equation}{eq:nse} with the lubrication theory under the condition of $ a/l\ll 1$. Under such condition, the second partial derivatives with respect to the axial coordiante $x$ are much less than those with respect to the radial coordiant $r$. Thus, at order $ (a/l)^0 $, we could neglect the axial derivatives, except the driving force terms $ \p_x p $, $ \p_x \psi $, $ \p_x \varphi $ and $ \p_x T $. Meanwhile, taking advantage of $ \p_r T\simeq 0 $ and $ u_r \simeq 0 $, \crefrange{nernst_planck_equation}{eq:nse} reduce to
\begin{equation}\label{eq:conc}
\frac{\p n_{\pm}}{\p r}\pm \frac{en_{\pm}}{k_{\mathrm{B}} T}\frac{\p\psi}{\p r}=0,
\end{equation}
\begin{equation}\label{eq:poisson}
\frac{1}{r}\frac{\p}{\p r}\left(r \frac{\p\psi}{\p r} \right) = -\frac{e(n_+-n_-)}{\epsilon},
\end{equation}
\begin{equation}\label{eq:v_r}
-\frac{\p p}{\p r}-\rho_e\frac{\p \psi}{\p r}=0,
\end{equation}
\begin{equation}\label{eq:v_x}
\frac{\mu}{r}\frac{\p}{\p r}\left(  r \frac{\p u_x}{\p r}\right) -\frac{\p p}{\p x}
-\rho_e\frac{\p \psi}{\p x}
-\rho_e\frac{\p \varphi}{\p x}
-\frac{\gamma\epsilon}{2}\left(\frac{\p \psi}{\p r} \right)^2\frac{\p T}{\p x}=0,
\end{equation}
where $ \gamma=\p_T \epsilon/\epsilon $. In above equations, the total electric potential $ \phi $ is assumed as a linear superposition of the EDL potential $\psi(r,x)$ and the induced electric potential $\varphi(x)$ \cite{Gross1968}, i.e., $ \phi(r,x)=\psi(r,x)+\varphi(x) $. Furthermore, in the derivation of \cref{eq:conc} the contribution of the ion convection is neglected (Appendix~\ref{appA}) and the zero normal flux on the capillary surface (i.e., $ \p_r n_{\pm}\pm (en_{\pm}/k_{\mathrm{B}} T) \p_r \psi=0 $) is employed.

Integrating \cref{eq:conc} in $r$ direction yields the Boltzmann distribution
\begin{equation}\label{eq:boltzmann}
n_{\pm}=n(x)\mathrm{e}^{\mp\Psi},
\end{equation}
where $ \Psi=e\psi/(k_{\mathrm{B}} T) $. The physical meaning of $ n(x) $ is interpreted as follows. In the absence of EDL overlap $ n(x) $ is straightforward the number concentration of either ion species in the capillary centerline; whereas for cases of EDL overlapping, $ n(x) $ represents the virtual number concentration of either ion species. It is worth pointing out the virtual quantities are defined in a virtual electroneutral solution that is in equilibrium with any capillary cross section \cite{peters2016}. Note that at the steady state, in the virtual electroneutral solution $ \bm{j}_{\pm}=0 $ and $ n(x)=(n_{+}+n_{-})/2 $. In analogous to the studies on the streaming potential \cite{masliyah2006}, the contribution of the ion convection to the ion virtual concentration can be neglected. Then it follows from $ \bm{j}_{+}/D_{+}+\bm{j}_{-}/D_{-}=0 $ that $ n(x) $ approximately satisfies the classic Soret equilibrium \cite{wurger2010,dietzel2017}
\begin{equation}\label{eq:Soret}
\frac{\p \ln n}{\p x}=-S_{\mathrm{T}} \frac{\p T}{\p x},
\end{equation}
where $ S_{\mathrm{T}}=(S_{\mathrm{T},+}+S_{\mathrm{T},-})/2 $ is the average intrinsic Soret coefficient of ions and its variation with temperature is neglected \cite{Rasuli2008} for relatively small temperature difference. Consider that the electrolyte in the reservoir is  maintained at the ambient temperature $ T_{\infty} $ and concentration $ n_{\infty} $, then one naturally has $ n|_{T_{\infty}}=n_{\infty} $ as a boundary condition for \cref{eq:Soret}. Subsequently, it derived from \cref{eq:Soret} that $ n=n_{\infty} \mathrm{e}^{S_{\mathrm{T}}(T_{\infty}-T)} $.

Substituting \cref{eq:boltzmann} into \cref{eq:poisson}, one finds
\begin{equation}\label{eq:pbe}
\frac{1}{\bar{r}}\frac{\p}{\p\bar{r}}\left(\bar{r}\frac{\p\Psi}{\p\bar{r}} \right)=\bar{\kappa}^2\sinh(\Psi),
\end{equation}
which is subject to a symmetry boundary condition at capillary axis and a constant $ \zeta $-potential condition on capillary surface, namely
\begin{equation}\label{eq:pb_bc}
\p_{\bar{r}}\Psi|_{\bar{r}=0}=0, \quad \Psi|_{\bar{r}=1}=\frac{e\zeta}{k_{\mathrm{B}} T} \equiv \bar{\zeta} ,
\end{equation}
where $ \bar{r}=r/a $ and
\begin{equation}\label{eq:local_debye}
\bar{\kappa} \equiv \kappa a =\bar{\kappa}_0\sqrt{\frac{n}{n_{\infty}}\frac{\epsilon_{\mathrm{ref}}}{\epsilon}\frac{T_{\mathrm{ref}}}{T}}
\end{equation}
with $ \bar{\kappa}_0=a\sqrt{2e^2n_{\infty} /(\epsilon_{\mathrm{ref}} k_{\mathrm{B}} {T_{\mathrm{ref}}})} $ being the dimensionless Debye parameter and subscript ref denoting physical properties taken at ${T_{\mathrm{ref}}} \simeq 298\;$K$ $. In addition, the present work neglects the temperature dependence of $ \zeta $ as most studies did \cite{Rasuli2008,wurger2010,dietzel2016,dietzel2017}.

To evaluate the expression for fluid velocity and cross-sectional average velocity (volumetric flow rate), the boundary conditions for \cref{eq:v_r,eq:v_x} are given as
\begin{equation}\label{eq:ns_bc}
\left. p \right|_{\psi=0}=p_0(x), \quad
\left. \p_r u_x \right|_{r=0}=0, \quad
\left. u_x \right|_{r=a}=0.
\end{equation}
The first condition requires that the pressure in the virtual bulk electroneutral region is equal to the virtual hydrostatic pressure $ p_0 $ \cite{peters2016}, i.e., the difference between pressure and excess osmotic pressure (i.e., $ k_{\mathrm{B}}T\sum_{i} (n_{i}-n) $); the last two require that the velocity distribution is symmetric with respect to capillary axis and the velocity is zero on capillary wall, respectively.

With the help of \cref{eq:conc,eq:poisson}, \cref{eq:v_r} can be rewritten as $ \p_r [p-k_{\mathrm{B}} T (n_++n_-)] $=0. Then integrating it along $ r $ coordinate with the first boundary condition of \cref{eq:ns_bc}, one obtains $ p=p_0(x)+2 n k_{\mathrm{B}} T [\cosh(\Psi)-1] $.
Subsequently, we rewrite \cref{eq:v_x} with the help of the expression for $ p $ as
\begin{align}\label{eq:v_xx}
\frac{\mu}{r}\frac{\p}{\p r}\left( r \frac{\p u_x}{\p r} \right) = & \frac{\p p_0}{\p x}+2  k_{\mathrm{B}} T [\cosh(\Psi)-1]\frac{\p  n }{\p x}\nonumber\\
& - 2 e n \sinh(\Psi) \frac{\p \varphi }{\p x} + 2n k_{\mathrm{B}}  \mathscr{H} \frac{\p T }{\p x},
\end{align}
where
\[
\mathscr{H}=\cosh(\Psi)-1-\Psi\cosh(\Psi) + \frac{\gamma^{\ast}}{2\bar{\kappa}^2}\left(\frac{\p \Psi}{\p \bar{r}} \right)^2,
\]
with $ \gamma^{\ast} = T\gamma $.

Then, we solve \cref{eq:v_xx} with the last two boundary conditions in \cref{eq:ns_bc} to obtain the axial velocity of the flow as
\begin{widetext}
\begin{align}\label{eq:u_x}
u_x = &-\frac{a^2 -r^2}{4 \mu}\frac{\p p_0}{\p x}-\frac{2 k_{\mathrm{B}}T}{\mu} \frac{\p  n}{\p x} \int_{r}^{a} \frac{\diff r_1}{r_1} \int_{0}^{r_1} r_2 [\cosh(\Psi)-1] \diff r_2  - \frac{\epsilon (\psi-\zeta)}{\mu}  \frac{\p \varphi}{\p x} -\frac{2nk_{\mathrm{B}}}{\mu} \frac{\p T}{\p x} \int_{r}^{a} \frac{\diff r_1}{r_1}\nonumber\\
& \times \int_{0}^{r_1} \left[ \cosh(\Psi)-1-\Psi\cosh(\Psi) + \frac{\gamma^{\ast}}{2\bar{\kappa}^2}\left(\frac{\p \Psi}{\p \bar{r}} \right)^2\right] r_2 \diff r_2,
\end{align}	
\end{widetext}
On the right-hand side of above equation, the first and third terms are well-known Hagen-Poiseuille flow and the electroosmotic flow, respectively; The second and last terms are essentially due to the osmotic pressure \cite{dietzel2017,Maheedhara2018} and the dielectric body force \cite{derjaguin1987} caused by the temperature dependence of the relative permittivity, and are termed TOF in the present work.

To obtain the cross-sectional average current density, we need the expression for the axial component of ion fluxes, which are derived from \cref{eq:density_flux} as
\begin{align}\label{eq:flux}
j_{\pm}=&-n_{\pm}D_{\pm}\left[\p_x \ln n\pm \frac{e \p_x \varphi}{k_{\mathrm{B}} T}+ \left( S_{\mathrm{T},\pm}\pm \frac{e\psi}{k_{\mathrm{B}} T^2} \right)\p_x T   \right]\nonumber\\
& +n_{\pm}u_x.
\end{align}

With the help of \cref{eq:u_x,eq:flux}, the average velocity, $ V=(2/a^2)\int_{0}^{a}u_x r \mathrm{d}r $, and the average current density, $ I=(2 e/a^2) \int_{0}^{a} (j_{+}-j_{-}) r \mathrm{d} r $, are derived as
\begin{align}
V = & \left( \frac{a^2}{8\mu}, \frac{n k_{\mathrm{B}} T  a^2}{\mu} L_1, \frac{\epsilon_{\mathrm{ref}} k_{\mathrm{B}} T_{\mathrm{ref}} }{\mu e}L_2, \frac{n k_{\mathrm{B}}  a^2}{\mu}L_3\right) \bm{X},\label{average_velocity}\\
I = & \left(\frac{\epsilon_{\mathrm{ref}} k_{\mathrm{B}} T_{\mathrm{ref}} }{\mu e}L_2, 4 (\beta L_4 + L_5) e n D , \frac{4 (\beta L_6+L_7) e^2 n D}{k_{\mathrm{B}} T}, \right. \nonumber\\
&  \left. \frac{4 (\beta L_8+S^{\dagger}_{\mathrm{T}}TL_9+L_{10}) e n D}{T} \right) \bm{X}, \label{eq:current}
\end{align}
where $\bm{X} =(-\partial_x p_0, -\partial_x \ln n, -\partial_x \varphi, -\partial_x T)^{\top} $ is the driving forces, $D=(D_++D_-)/2$ is the average diffusion coefficient of ions, $ S^{\dagger}_{\mathrm{T}} = (D_+S_{\mathrm{T},+}+D_-S_{\mathrm{T},-})/(D_++D_-) $ is the diffusivity-modified Soret coefficient of ions and $ \beta = (\epsilon_{\mathrm{ref}} / \mu_{\mathrm{ref}} D_{\mathrm{ref}}) (k_{\mathrm{B}}{T_{\mathrm{ref}}}/e)^2 $ is the so-called intrinsic P{\'e}clet number \cite{saville77}. The coefficients $ L_i $ in \cref{average_velocity,eq:current} are given as follows:
\begin{align}\label{eq:Lmat}
&L_1=\int_0^1(\bar{r}-\bar{r}^3) [\cosh(\Psi)-1]\diff \bar{r}\nonumber\\
&L_2=2\frac{\epsilon}{\epsilon_{\mathrm{ref}}}\int_{0}^{1} (\bar{\psi}-\bar{\zeta})\bar{r}\diff \bar{r}\nonumber\\
&L_3=\int_{0}^{1} (\bar{r}-\bar{r}^3) \mathscr{H}\diff \bar{r}\nonumber\\
&L_4=\frac{\epsilon}{\epsilon_{\mathrm{ref}}}\int_0^1 (\bar{\psi}-\bar{\zeta}) [\cosh(\Psi)-1]\bar{r}\diff \bar{r}\nonumber\\
&L_5=\int_0^1 [\chi_{D}\cosh(\Psi)-\sinh(\Psi)]\bar{r}\diff \bar{r}\nonumber\\
&L_6=-\frac{\epsilon}{\epsilon_{\mathrm{ref}}}\int_{0}^{1} \bar{r}(\bar{\psi}-\bar{\zeta})\sinh(\Psi)\diff \bar{r}\nonumber\\
&L_7=\int_0^1 [\cosh(\Psi)-\chi_{D}\sinh(\Psi)]\bar{r}\diff \bar{r}\nonumber\\
&L_8=\frac{\epsilon}{\epsilon_{\mathrm{ref}}}\int_0^1 (\bar{\psi}-\bar{\zeta}) \mathscr{H}\bar{r} \diff \bar{r}\nonumber\\
&L_9=\int_0^1 [\chi_{Q}^{\dagger}\cosh(\Psi)-\sinh(\Psi)]\bar{r}\diff \bar{r}\nonumber\\
&L_{10}=\int_0^1\Psi [\cosh(\Psi)-\chi_{D}\sinh(\Psi)]\bar{r}\diff \bar{r}
\end{align}
where $ \bar{\psi}=e\psi/(k_{\mathrm{B}}T_{\mathrm{ref}}) $, $ \chi_{D}=(D_+-D_-)/(D_++D_-) $, $ \chi_{Q}^{\dagger}=(D_+S_{\mathrm{T},+}-D_-S_{\mathrm{T},-})/(2S^{\dagger}_{\mathrm{T}}) $. Notably, in the course of derivation of \cref{eq:current}, we reduce the triple integrals in the convection current to single integrals by interchanging the order of integration.

At the steady state, the average current density $ I $ vanishes and a stable induced electric field is then built up. In this case, it follows from \cref{eq:current} that
\begin{equation}\label{eq:varphi}
-\frac{\p \bar{\varphi}}{\p x}
=A_0 \left( \frac{1}{4 \bar{n}} \frac{\partial \bar{p}}{\partial x} \right)
+A_1 \left( \bar{T} \frac{\partial \ln \bar{n}}{\partial x} \right)
+A_2 \left( \frac{\partial \bar{T}}{\partial x} \right),
\end{equation}
where $\bar{\varphi} = e\psi/(k_B T) $, $\bar{p}_0 = p_0/(n_{\infty} k_B T_{\mathrm{ref}}) $, $\bar{n} = n/n_{\infty}$, $\bar{T}= T/T_{\infty}$ and
\begin{align}\label{eq:coefficient}
& A_0=\frac{\beta L_2}{\beta L_6+L_7}, \quad
  A_1=\frac{\beta L_4 + L_5}{\beta L_6+L_7},\nonumber\\
& A_2= \frac{\beta L_8 + S^{\dagger}_{\mathrm{T}} T L_9 +L_{10}}{\beta L_6+L_7}.
\end{align}

Substituting \cref{eq:varphi} into \cref{average_velocity}, we obtain
\begin{equation}\label{eq:dpdx}
\frac{\p \bar{p}_0}{\p \bar{x}}
= B_0 \left( \frac{\mu}{\mu_{\mathrm{ref}}}\bar{V}\right)
+ B_1 \left( -\bar{T} \frac{\p \ln \bar{n}}{\p \bar{x}} \right) 
+ B_2 \left( -\frac{\p \bar{T}}{\p \bar{x}}\right),
\end{equation}
with $ \bar{V}=V/u_{\mathrm{ref}} $, $ \bar{x}=x/l $, and
\begin{align}\label{eq:coefficient_b}
& B_0=\frac{4\lambda^2}{\lambda^2 A_0 L_2 n_{\infty}/n-1/4}, \nonumber\\
& B_1=B_0 \left( A_1 L_2 -\frac{L_1}{2\lambda^2} \frac{n}{n_{\infty}}  \right), \nonumber  \\
& B_2=B_0\left( A_2 L_2 -\frac{L_3}{2\lambda^2} \frac{n}{n_{\infty}}  \right),
\end{align}
where $ \lambda=\bar{\kappa}_0^{-1} $ is the dimensionless Debye length and $ u_{\mathrm{ref}}=(\epsilon_{\mathrm{ref}}/\mu_{\mathrm{ref}} l)(k_{\mathrm{B}}T_{\mathrm{ref}}/e)^2 $ is the reference velocity.

Subsequently, integrating both side of \cref{eq:dpdx} with respect to $ x $-coordinate, keeping average velocity $ \bar{V} $ constant and simultaneously noting that $ \Delta p_0 =0 $, we have
\begin{equation}\label{calc_average_velocity}
\bar{V} = \frac{\int_0^1 B_1 \bar{T} \p_{\bar{x}}\ln n \diff \bar{x} +\int_0^1 B_2 \p_{\bar{x}} \bar{T} \diff \bar{x} }{\int_0^1 \bar{\mu}  B_0 \diff \bar{x}}.
\end{equation}
where $ \bar{\mu} = \mu/\mu_{\mathrm{ref}}$.

Once the average velocity is determined, the thermoelectric potential can be integrated from \cref{eq:varphi} as
\begin{align}\label{eq:varphi_diff}
-\Delta\bar{\varphi} = 
-\frac{e \Delta \varphi}{k_{\mathrm{B}} {T_{\mathrm{ref}}}}
& = \bar{V}\int_{0}^{1}\frac{\bar{\mu}}{\bar{n}}\frac{A_0 B_0}{4} \diff\bar{x} \nonumber\\
& - \int_{0}^{1} \frac{A_0 (B_2  -S_{\mathrm{T}}T B_1)}{4\bar{n}}  \frac{\p \bar{T}}{\p \bar{x}} \diff \bar{x} \nonumber\\
& + \int_0^1 (A_2 - S_{\mathrm{T}}T A_1) \frac{\p \bar{T}}{\p \bar{x}} \diff \bar{x},
\end{align}
in which the first two terms account for the axial variation of $ p_0 $ and would be negligibly small if $ \Delta p_0 = 0 $ although there exists a perturbation of the local induced electric field $ -\p \varphi/\p x $ due to non-zero $ \p p_0/\p x $ in the capillary, and the last term is main part of the induced electric potential, which is the focus of this paper. Neglecting the first two terms on the right-hand side of \cref{eq:varphi_diff}, with some mathematical operations one obtains
\begin{align}\label{eq:main_potential}
S_e = -\frac{\Delta\varphi}{\Delta T} 
& \approx \frac{k_B}{e \Delta T} \int_{T_{\infty}}^{T_{\infty}+\Delta T} (S_{\mathrm{T}} T C_0 + C_1) \mathrm{d}T \nonumber\\
& + \frac{k_B}{e \Delta T} \int_{T_{\infty}}^{T_{\infty}+\Delta T} C_2 \mathrm{d}T \nonumber\\
& + \frac{k_B}{e \Delta T} \int_{T_{\infty}}^{T_{\infty}+\Delta T} \chi_{Q} S_{\mathrm{T}} T C_3 \mathrm{d}T,
\end{align}
where $ \chi_{Q}=(S_{\mathrm{T},+}-S_{\mathrm{T},-})/(2S_{\mathrm{T}}) $ and
\begin{align}\label{eq:coefficient_c}
& C_0=  -\frac{\beta L_4}{\beta L_6+L_7},\quad 
C_1=   \frac{\beta L_8}{\beta L_6+L_7},\nonumber\\
& C_2=   \frac{L_{10}}{\beta L_6+L_7}, \quad
C_3=   \frac{L_{7}}{\beta L_6+L_7}.
\end{align}
It is worth explaining the physical meanings of the terms on the right-hand side of \cref{eq:main_potential}: (a) The first terms measures the thermoelectric effect due to the TOF. Practically, the TOF can drive the ion charges in the EDL to the downstream and a thermoosmotic streaming current forms accordingly. For the open-circuit system and the steady state, the accumulation of charges in the downstream can produce a reverse electric field to balance the streaming current \cite{dietzel2017}. This mechanism is referred to as thermoelectric effect A (TE-A) in the present work. (b) The second term represents the thermoelectric effect due to the temperature dependence of ion electrophoretic mobilities. Such temperature dependence of ion electrophoretic mobilities can result in axial ion concentration gradients in the EDL, which cause charge separation and thus produce an electric field \cite{dietzel2016}. This mechanism is referred to as thermoelectric effect B (TE-B) in the present work. (c) The third term represents the thermoelectric effect due to the difference in the intrinsic Soret coefficient of cation and anion. In analogous to the Seebeck effect in (semi-)conductors, such difference also can lead to charge separation and thus establishes an electric field. This well-accepted mechanism is referred to as thermoelectric effect C (TE-C) in the present work.

Finally, we briefly describe the numerical calculation of the semi-analytical model developed above. First, the axial domain $ 0 \leq \bar{x} \leq 1 $ was partitioned into $ N $ equal intervals, and then \cref{eq:pbe,eq:pb_bc} were numerically solved by the finite element method (COMSOL Multiphysics) at discrete points $ \bar{x}_i=i/N $ ($ i=0,\cdots,N $). To check the grid independence, we compare the calculated potentials at $ \bar{r}=0 $ and $ \bar{r} \simeq 0.9 $ under different grid numbers. The grid-independence study shows that it is reasonable to employ a mesh of 1000 elements in $ \bar{r} $ direction with an element ratio of at least 10 (to refine the mesh near the capillary wall). Subsequently, coefficients $ L_j $ in \cref{eq:coefficient} were calculated as a function of $ T(\bar{x}_i)=T_{\infty}+\bar{x}_i\Delta T $. Then, the average velocity $ \bar{V} $ and the thermoelectric potential $ \Delta\varphi $ were numerically calculated from \cref{calc_average_velocity,eq:varphi_diff}, respectively. To calculate the axial velocity, we determined the local pressure gradient and local electric field from Eqs.~(\ref{eq:dpdx}) and (\ref{eq:varphi}), respectively. Then, the velocity was calculated from \cref{eq:u_x}. All numerical integrations in above calculations were evaluated by \texttt{trapz} function in MATLAB. For the calculation of $ L_j $ in \cref{eq:coefficient}, the number of grid points is determined by that used for solving \cref{eq:pbe,eq:pb_bc}. For integrals in $ \bar{x} $-direction, the number of grid points is equal to $ N+1 $ (here $ N=100 $). We confirm the grid independence of the present meshes by comparing the variations of the calculated values of these integrals before and after refining the grid.

\section{Results and discussion}\label{results}

The physical properties for three widely used alkali chloride electrolytes as well as the temperature dependence of the viscosity and of the permittivity are summarized in Appendix~\ref{appB}, and the validation of the semi-analytical model is given in Appendix~\ref{appC}. Additionally, in our calculations the ambient temperature and temperature difference are respectively set to $ T_{\infty}/{T_{\mathrm{ref}}}=1 $ and $ \Delta T/{T_{\mathrm{ref}}}=0.0671 $ (e.g., $ \Delta T=20 $ K), unless otherwise stated.

\subsection{Analysis of thermoelectric effects}\label{sec:te_effect}

\begin{figure*}
\centering{\includegraphics[width=0.75\linewidth]{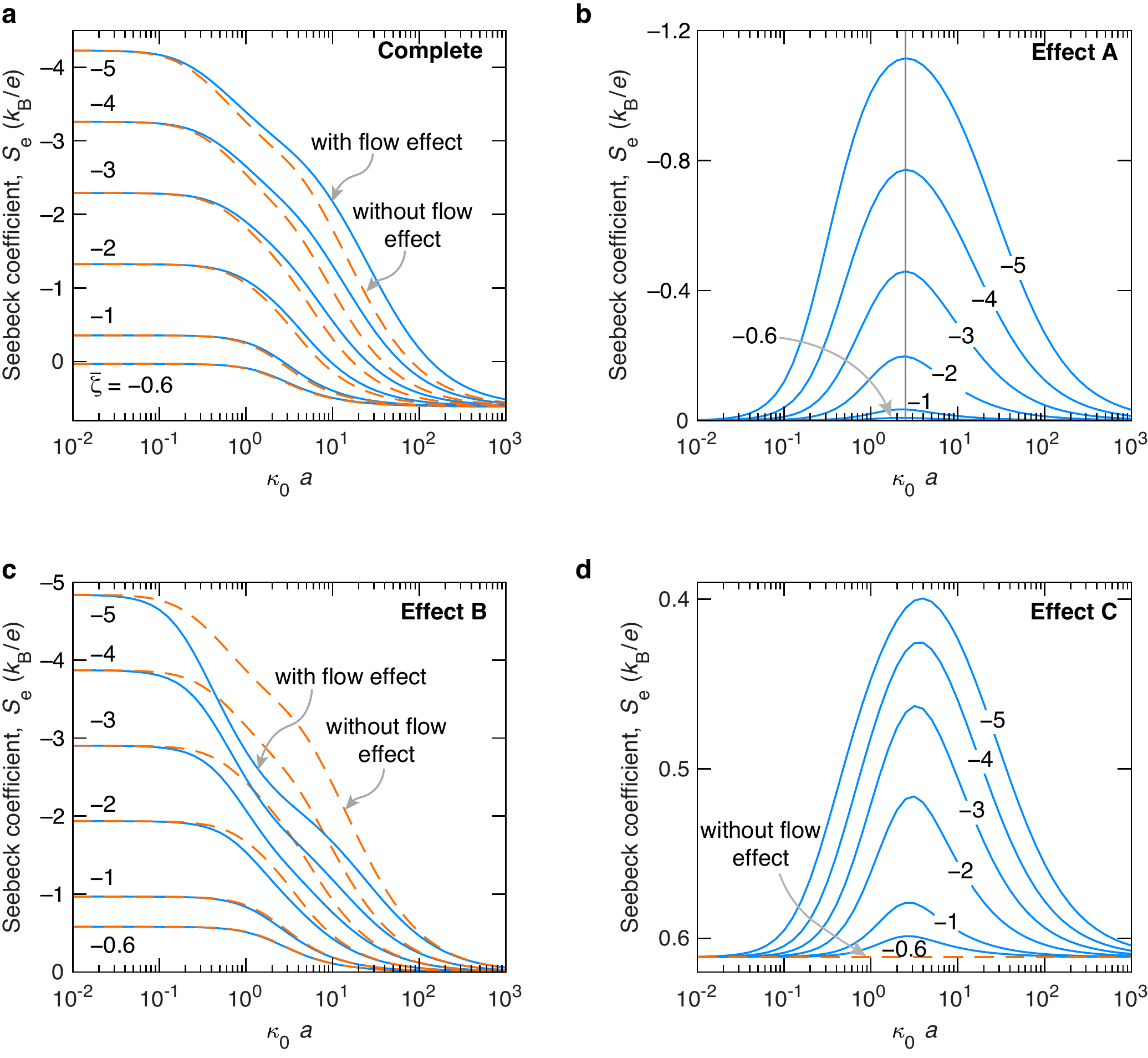}}
\caption{\label{fig:Seebeck}Seebeck coefficient $ S_e $ (in units of $ k_{\mathrm{B}}/e $) as a function of dimensionless Debye parameter $ \bar{\kappa}_0 $ for an aqueous NaCl solution with $ \bar{\zeta}=e\zeta/(k_{\mathrm{B}} T_{\mathrm{ref}})=-0.6, -1,-2,-3,-4,-5 $. The ambient temperature and temperature difference are set to $ T_{\infty}/{T_{\mathrm{ref}}}=1 $ and $ \Delta T/{T_{\mathrm{ref}}}=0.0671 $, respectively. (a) Comparison between complete solutions (solid lines) and solutions of the hypothetical case of no liquid flow ($ \beta=0 $, dashed lines). (b) Thermoelectric effect A. (c) Thermoelectric effect B for $ \beta \neq 0 $ (solid lines) and $ \beta=0 $ (dashed lines). (d) Thermoelectric effect C for $ \beta \neq 0 $ (solid lines) and $ \beta=0 $ (dashed lines).}	
\end{figure*}

The thermoelectric effect of electrolyte solutions can be characterized by the Seebeck coefficient $ S_e $, which is the focus of this section. \cref{fig:Seebeck}a shows $ S_e $ for \ce{NaCl} solution as a function of $ \bar{\kappa}_0 $ with varying $ \bar{\zeta} = e\zeta/(k_{\mathrm{B}} T_{\mathrm{ref}})$. To check the effect of fluid flow, results for hypothetical case of no fluid flow ($ \beta=0 $) are also included in this plot. For $ \bar{\zeta}=-0.6 $ (i.e., $ \zeta \simeq -15 $ mV), the Seebeck coefficient $ S_e $ is always positive, namely the thermoelectric field directs towards the hot end, and the magnitude of $ S_e $ increases with increasing  $ \bar{\kappa}_0 $. As $ \bar{\kappa}_0 \tti $, $ S_e $ goes to a constant ($ S_e^{\infty}\simeq 0.61 k_{\mathrm{B}}/e $) and so is $ \Delta \varphi $ ($ =-S_e^{\infty}\Delta T $; see Appendix~\ref{appC}). By contrast, for higher $ \zeta $ potentials, $ S_e $ becomes negative under the extreme confinement (i.e., $ \bar{\kappa}_0 \ttz $) and increases to zero as $ \bar{\kappa}_0 $ increases to a critical value $ \bar{\kappa}_{0,cr}$ (e.g., $ \bar{\kappa}_{0,cr} \simeq [2.7,\, 9.5,\, 25.1,\, 60.9,\, 137.1] $ for $ \bar{\zeta}=[-1,\, -2,\,-3,\,-4,\,-5] $, see \cref{fig:Seebeck}a). With $ \bar{\kappa}_0 $ exceeding $ \bar{\kappa}_{0,cr}$, the sign of $ S_e $ switches from negative to positive,~i.e., the direction of the thermoelectric field is reversed. If one continues to increase $ \bar{\kappa}_0 $, $ S_e $ is expected to increase until reaching ca. $ 0.61k_{\mathrm{B}}/e $ at $ \bar{\kappa}_0 \tti $. It is also clear that the critical $ \bar{\kappa}_{0,cr}$ of sign reversal increases as $ |\bar{\zeta}| $ increases.

Clearly, for $ |\bar{\zeta}| \lesssim 1 $ the fluid flow has a negligible effect on $ S_e $, while for larger $ |\bar{\zeta}| $ the fluid flow influences substantially $ S_e $ in a certain range of $  \bar{\kappa}_0 $ (\cref{fig:Seebeck}a), e.g., for $ \bar{\zeta} =-3 $ in the range of $ 10\lesssim\bar{\kappa}_0\lesssim 50 $, the Seebeck coefficient excluding fluid flow effect can deviate from the complete solution by more than 10\%. The higher the $ |\bar{\zeta}| $ value is, the more significantly the fluid flow influences the Seebeck coefficient. However, under the two extreme conditions, i.e., $ \bar{\kappa}_0 \tti $ or $ \bar{\kappa}_0 \ttz $, regardless of $ \bar{\zeta} $ values, the effect of the fluid flow on $ S_e $ disappears. In fact, from \cref{eq:coefficient_c} it shows that the fluid flow gives rise to two effects: (\romannumeral1) inducing a thermoosmotic streaming current \cite{dietzel2017} measured by $ L_4 $ and $ L_8 $ and (\romannumeral2) increasing the electric conductivity of solutions measured by $ L_6 $. Both these two effects vanish under aforementioned two extreme conditions. Finally, one concludes that the assumption of no liquid flow is generally valid for very small and very large $\bar{\kappa}_0$, but is not valid for a finite $\bar{\kappa}_0$ especially when $ |\bar{\zeta}| $ values are high.

To have a more detailed understanding of the characteristics shown in \cref{fig:Seebeck}a, the individual contributions to the overall thermoelectric effect due to the TE-A, TE-B and TE-C are shown in \cref{fig:Seebeck}b--d, respectively. Evidently, the Seebeck coefficient associated with the TE-A, $ S_{e}^{\mathrm{A}} $, is always negative (\cref{fig:Seebeck}b), which indicates that the corresponding thermoelectric field is directed towards the cold end. Furthermore, the magnitude of $ S_{e}^{\mathrm{A}} $ increases with increasing $ |\bar{\zeta}| $. The thermoelectric potential due to the TE-A shares the same mechanism as the pressure-driven streaming potential, both being caused by the charge separation due to the flow-induced streaming current. Nevertheless, in contrast to the conventional streaming potential which arrives at the maximum value as $ \bar{\kappa}_0 \tti $, the thermoelectric potential here reaches its maximum value at $ \bar{\kappa}_0\simeq 2.5 $ (in comparison, $ \bar{\kappa}_0\simeq 2 $ for slit channels \cite{dietzel2017}), and asymptotically reduces to zero as $ \bar{\kappa}_0 \ttz $ or $ \bar{\kappa}_0 \tti $. The reason for this behavior has been detailed in ref.~\cite{dietzel2017}. Quantitatively, for $ \bar{\zeta}=-5 $ the maximum value of $ | S_{e}^{\mathrm{A}} | $ is equal to ca. $ 1.11k_{\mathrm{B}}/e $ ($ \simeq 96\;\mathrm{\mu}$V$\;$K$^{-1} $), which amounts to 37.4\% of the $ S_e $ at $ \bar{\kappa}_0 \simeq 2.5 $. Notably, such percentage is much higher than the effect of the fluid flow on $ S_e $ (\cref{fig:Seebeck}a), which is due to the increase of the electric conductivity caused by the EOF.

The TE-B was discovered theoretically and has been demonstrated as a confinement effect \cite{dietzel2016}. For each value of $ \bar{\zeta} $ ($ <0 $), the magnitude of the Seebeck coefficient associated with the TE-B, $ S_{e}^{\mathrm{B}} $, decreases from a maximum value to zero as $ \bar{\kappa}_0 $ increases from zero to infinity and is also an increasing function of $ |\bar{\zeta}| $ (\cref{fig:Seebeck}c). For instance, with the value of $ |\bar{\zeta}| $ increasing from $ 1 $ to $ 5 $, the maximum value of $ |S_{e}^{\mathrm{B}}| $ increases from $ 0.97k_{\mathrm{B}}/e $ to $ 4.8k_{\mathrm{B}}/e $ (ca. $ 83 $--$ 417 \;\mathrm{\mu}$V$\;$K$^{-1} $). We find that the TE-B is substantially weaken by the presence of the fluid flow, especially for moderate values of $ \bar{\kappa}_0 $. For instance, for $ \bar{\zeta}=-5 $ and $ \bar{\kappa}_0\simeq 4 $ the magnitude of $ S_{e}^{\mathrm{B}} $ even declines by more than 30\% because of the EOF. Furthermore, the non-positive values of $ S_e $ implies that in our calculation case, the direction of the thermoelectric field due to the TE-B is directed to the cold end, which is consistent with the direction of the thermoelectric field due to the TE-A.

The Seebeck coefficient affiliated with the TE-C, $ S_{e}^{\mathrm{C}} $, tends to a $ \zeta $-independent value ($ S_e^{\infty} \simeq 0.61 k_{\mathrm{B}}/e $) as $ \bar{\kappa}_0 \ttz $ or $ \bar{\kappa}_0 \tti $ (\cref{fig:Seebeck}d), indicating that in these two limiting cases the TE-C is unaffected by the EDL. By contrast, the magnitude of $ S_{e}^{\mathrm{C}} $ is found to be lowered by the presence of the EDL for a moderate value of $ \bar{\kappa}_0 $. This trend is enhanced by increasing the value of $ |\bar{\zeta}| $. For instance, for $ \bar{\zeta}=-0.6 $ the minimum value of $ S_{e}^{\mathrm{C}} $ is ca. $ 0.98S_e^{\infty} $ (obtained at $ \bar{\kappa}_0 \simeq 2.5 $), whereas for $ \bar{\zeta}=-5 $ such minimum value becomes only ca. $ 0.65S_e^{\infty} $ (obtained at $ \bar{\kappa}_0 \simeq 4 $). The reason for the reduction of the magnitude of $ S_{e}^{\mathrm{C}} $ lies in the increase of the electric conductivity due to the EOF induced by the thermoelectric field. Furthermore, it is noted that in our calculation case, the $  S_{e}^{\mathrm{C}}  $ is positive, which means that the corresponding thermoelectric field is directed to the hot end. Clearly, the direction of the thermoelectric field due to the TE-C is opposite to that of the thermoelectric fields due to the TE-A and TE-B.

Based on the characteristics of three individual $ S_e $ in \cref{fig:Seebeck}b--d, the reason for the sign reversal of $ S_e $ in \cref{fig:Seebeck}a is now straightforward to understand.


So far, we have shown that for $ S_{\mathrm{T}}>0$ (thermophobic electrolytes), with $ \bar{\zeta}<0 $ the thermoelectric fields due to the TE-A and TE-B are counteracted by the thermoelectric filed due to the TE-C, this normally leads to a reduction of $ |S_e| $. Naturally, we may propose such a question: does there exist a condition for the the TE-C to cooperate with the the TE-A and TE-B to maximize $ |S_e| $? The analysis of our analytical results (\cref{eq:main_potential}) give an affirmative answer to this question: for $ S_{\mathrm{T}}>0 $ the three types of thermoelectric effects are found to be in full cooperation as long as $  \zeta\chi_{Q} S_{\mathrm{T}} >0 $, while for $ S_{\mathrm{T}}<0 $ (thermophilic electrolytes), other than the above condition a supplemental condition of $ |C_0/C_1|_{\mathrm{ave}}>|S_{\mathrm{T}} T |_{\mathrm{ave}} $ is required, where subscript ave represents average temperature.

\begin{figure}[!htb]
	\centering
	\includegraphics[width=1\linewidth]{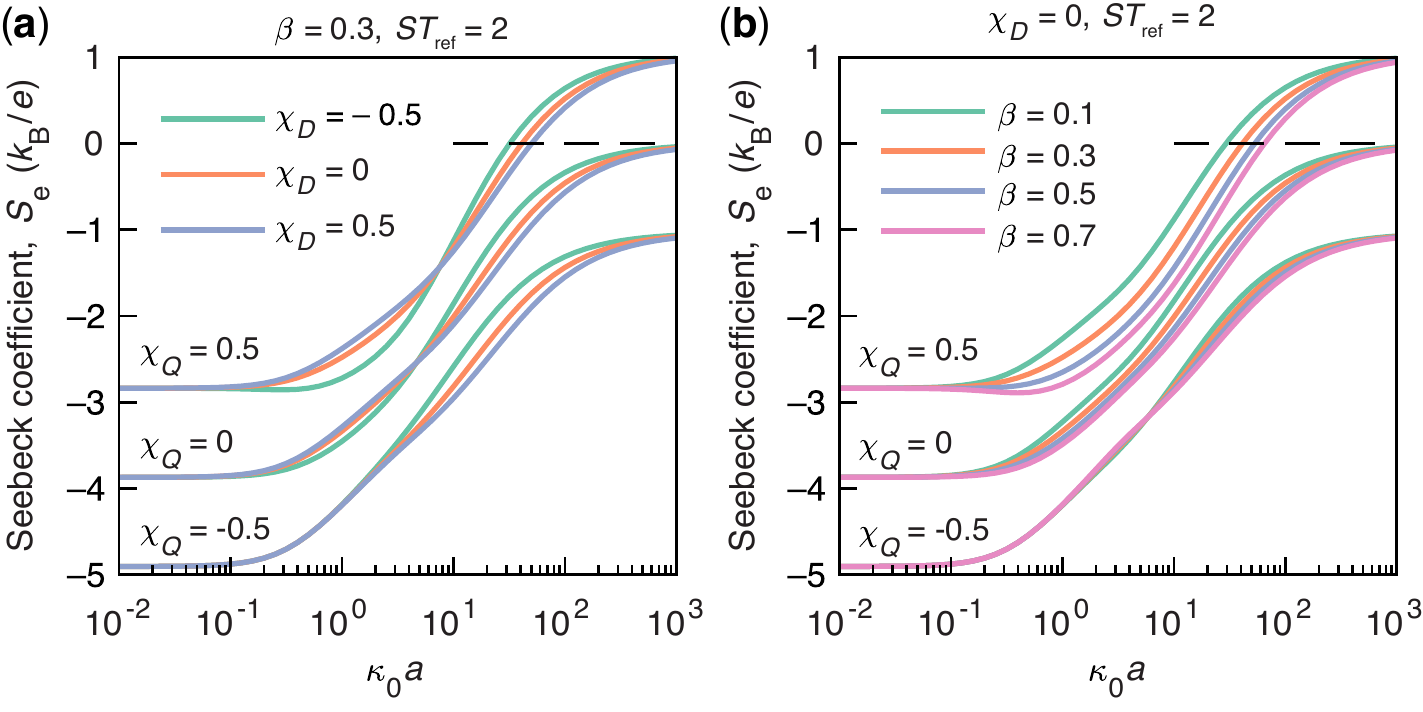}
	\caption{Seebeck coefficient $ S_e $ (in units of $k_B/e$) as a function of dimensionless Debye parameter $ \bar{\kappa}_0 $ for (a) varying $ \chi_{D} $ and $ \chi_{Q} $ at $ \beta=0.3 $, $ S_{\mathrm{T}}{T_{\mathrm{ref}}}=2 $ and (b) varying $ \beta $ and $ \chi_{Q} $ at $ \chi_{D}=0 $, $ S_{\mathrm{T}}{T_{\mathrm{ref}}}=2 $. The $ \zeta $ potential is set to $ \bar{\zeta}=-4 $. The temperature conditions are the same as those in \cref{fig:Seebeck}.}
	\label{fig:parametric_Se}
\end{figure}

Next, we investigate the effects of electrolytes on the Seebeck coefficient.
It follows from \cref{eq:main_potential} that there are four variables which distinguish one type of electrolyte from another, i.e., $ \chi_{Q} $, $\chi_{D}$, $ S_{\mathrm{T}}{T_{\mathrm{ref}}} $ and $ \beta $. The first two characterize the differences in intrinsic Soret coefficients and diffusion coefficients between cations and anions, respectively. The last two characterize the average values of intrinsic Soret coefficients and diffusion coefficients for cations and anions, respectively. Thus the effects of electrolytes on $ S_e $ can be reflected by the above four variables.
Evidently, neither $ \chi_{D} $ nor $ \beta $ has an effect on $ S_e $ as $ \bar{\kappa}_0\ttz $ or $ \bar{\kappa}_0\tti $ (\cref{fig:parametric_Se}). For $\bar{\kappa}_0\tti$, the thermoelectric potential approaches the $ \chi_{D} $- and $ Pe $-independent bulk Soret voltage (see Appendix~\ref{appC}). For $ \bar{\kappa}_0\ttz $, the difference in the ion diffusivities is of no importance since the thermoelectric effect is primarily caused by the counter-ions, and $ \beta $ also does not play a role in $ S_e $ because of the vanishing fluid flow.

For moderate $ \bar{\kappa}_0 $ values, both $ \chi_{D} $ and $ \beta $ have substantial effects on $ S_e $. \cref{fig:parametric_Se}a shows that for $ \chi_{Q}=0 $, there exist a transition point for $ \bar{\kappa}_0 $ ($ \simeq 5 $) where the values of $ S_e $ for different $ \chi_{D} $ are almost equal. If $ \bar{\kappa}_0 $ is below (above) such transition point, an increase of $ \chi_{D} $ can lead to a higher (lower) value of $ S_e $. Such transition point for $  \bar{\kappa}_0 $ normally varies with $ \chi_{Q} $. For example, if $ \chi_{Q}=0.5 $ such transition point for $  \bar{\kappa}_0 $ increases to ca. 8, while almost disappears if $ \chi_{Q}=-0.5 $. Notably, the transition point for  $ \bar{\kappa}_0 $ can not be observed for smaller values of $ S_{\mathrm{T}}T_{\mathrm{ref}} $ (e.g., $ S_{\mathrm{T}}T_{\mathrm{ref}}=1 $; not shown). The characteristics shown in \cref{fig:parametric_Se}a can be explained as follows. For $ \bar{\zeta}<0 $, an increase of $ \chi_{D} $ implies that the diffusivity of cations (dominant charge carriers in the EDL) becomes larger and thus gives rise to following two effects: (\romannumeral1) the electric conductivity measured by $ \beta L_6+L_7 $ increases and thus there is a tendency to reduce $ |\Delta \varphi| $; (\romannumeral2) the current affiliated with either the TE-B or the TE-C (measured by $ L_{10 }$ or $ L_7 $, respectively) increases, which can result in a larger $ |\Delta \varphi| $. For $ \chi_{Q}=0 $ (the TE-C vanishes), the effects \romannumeral1~and \romannumeral2~counteract each other and the effect \romannumeral1~(\romannumeral2) is dominant if $ \bar{\kappa}_0 $ is below (above) the corresponding transition point. As $ \chi_{Q} $ deviates from 0, the effect \romannumeral1~remains unchanged, while the effect~\romannumeral2~becomes weaker ($ \chi_{Q}>0 $)  or  stronger ($ \chi_{Q}<0 $) due to the counteraction or cooperation of TE-B and TE-C, respectively. Eventually, the variation in effect \romannumeral2~leads to the change of the transition point for $ \bar{\kappa}_0 $.

\cref{fig:parametric_Se}b reveals that for $ \chi_{Q}=0 $, the Seebeck coefficient $ S_e $ (with sign) decreases with increasing $ \beta $. Such behavior becomes more (less) pronounced when increasing (decreasing) $ \chi_{Q} $. For $ \chi_{Q}=-0.5 $ with $ 1 \lesssim \bar{\kappa}_0 \lesssim 5 $, one even finds that $ S_e $ increases with the increase of $ \beta $. Practically, increasing $ \beta $ on one hand increases the efficient electric conductivity due to the EOF \cite{Siria2013} by noting that $\beta \propto $ the ion diffusion coefficient, on the other hand increases the thermoosmotic streaming current. In analogy to previous analysis of the influence of $ \chi_{D} $ on $ S_e $, the behavior shown in \cref{fig:parametric_Se}b originates from the competition between these two effects (note that an electric potential is the ratio of an electric current to an electric conductance).

\begin{figure*}[!htb]
	\centering
	\includegraphics[width=0.75\linewidth]{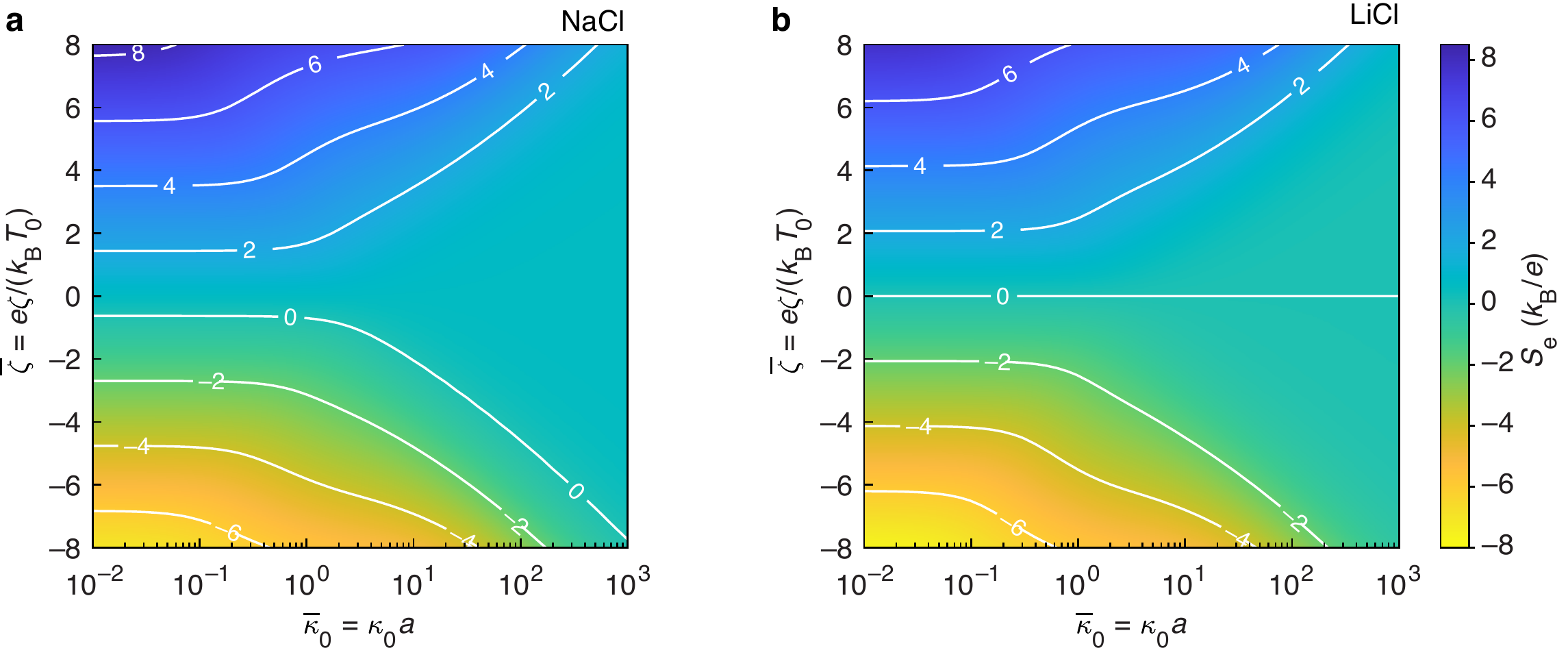}
	\caption{Contour plots of Seebeck coefficient $ S_e $ (in units of $ k_{\mathrm{B}}/e $) for (a) NaCl and (b) LiCl in a $ \bar{\kappa}_0 $--$ \bar{\zeta} $ domain. The temperature conditions are the same as those in \cref{fig:Seebeck}.}
	\label{contour1}
\end{figure*}

Next, to give a comprehensive view on the effects of $ \bar{\kappa}_0 $ and $ \bar{\zeta} $ on $ S_e $ for various practical electrolytes with electrolyte-specific parameters, in \cref{contour1}a and b we show contour plots of $ S_e $ as a function of $ \bar{\kappa}_0 $ and $ \bar{\zeta} $ for~\ce{NaCl} and~\ce{LiCl}, respectively. In the~\ce{NaCl} solution, for positive or lowly negative $ \bar{\zeta} $, $ S_e>0 $ regardless of $ \bar{\kappa}_0 $ values, while for highly negative $ \bar{\zeta} $, $  S_e>0 $ is only present if $ \bar{\kappa}_0 $ exceeds the corresponding critical $ \bar{\kappa}_{0,cr} $ of sign reversal (\cref{contour1}a). This means that $ S_e $ in the~\ce{NaCl} solution is asymmetric with respect to zero-$ \bar{\zeta} $ line due to non-zero $ \chi_{Q} $. On the contrary, in the~\ce{LiCl} solution, the contour plot of $  S_e  $ is nearly symmetric with respect to the line of $ \bar{\zeta}=0 $ (\cref{contour1}b) due to vanishing $ \chi_{Q} $, and accordingly there is no critical $ \bar{\kappa}_{0,cr} $ of sign reversal.


\begin{figure}[!htb]
	\centering
	\includegraphics[width=1\linewidth]{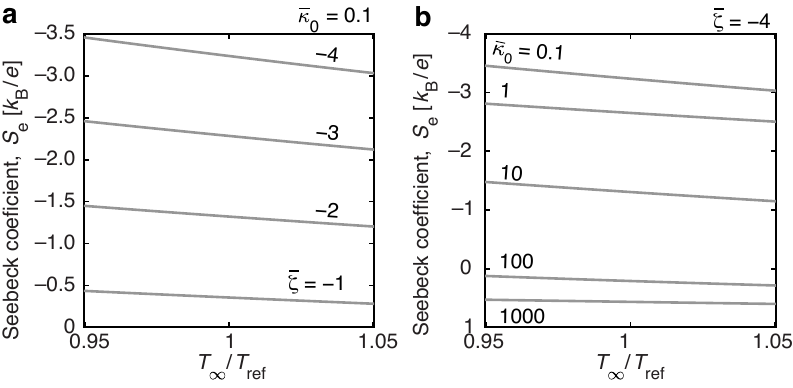}
	\caption{Effect of ambient temperature $ T_{\infty} $ on Seebeck coefficient $S_e $ (in units of $k_B/e$) of aqueous NaCl solution for (a) varying $ \bar{\zeta} $ as $ \bar{\kappa}_0=0.1 $, and (b) varying $ \bar{\kappa}_0 $ as $ \bar{\zeta}=-4 $. The temperature difference is set to $ \Delta T/T_{\mathrm{ref}}=0.0671 $. The lines are guide for the eye.}
	\label{fig:amb_temp}
\end{figure}

Then, we analyze the effect of the ambient temperature on the Seebeck coefficient in \ce{NaCl} solutions. Obviously, for $ \bar{\kappa}_0=0.1 $ and all chosen $ \bar{\zeta} $ ($= -1$, $ -2 $, $-3 $ or $-4 $), the magnitude of $ S_e $ nearly linearly decreases with increasing $ T_{\infty}/T_{\mathrm{ref}} $ (\cref{fig:amb_temp}a), Also, as $|\bar{\zeta}| $ increases $ |S_e| $ decreases with $ T_{\infty}/T_{\mathrm{ref}}  $ in an increasing rate. Similarly, one observes $ S_e $ is a nearly linear function of $ T_{\infty} $ for different values of $ \bar{\kappa}_0 $ ($ =0.1$--$ 1000 $) with $ \bar{\zeta}=-4 $ (\cref{fig:amb_temp}b). The corresponding slope (with sign) is found to be a increasing function of $ \bar{\kappa}_0 $. The characteristics shown in \cref{fig:amb_temp} can be explained as follows. For $ \bar{\kappa}_0\tti $, $ S_e = S_e^{\infty}=\chi_{Q} k_{\mathrm{B}} S_{\mathrm{T}} T_{\mathrm{ave}} /e $, whereas for $ \bar{\kappa}_0\ll 1 $ (e.g., $ \bar{\kappa}_0= 0.1 $), $ S_e \approx S_e^{\infty}+(\zeta/\Delta T)\ln(1+\Delta T/T_{\infty}) $ (Appendix~\ref{appC}), where $ T_{\mathrm{ave}} $ is the average temperature. The trends shown in \cref{fig:amb_temp} can be straightforwardly understood by analyzing $ S_e $ and its derivative with respect to $ T_{\infty}/T_{\mathrm{ref}} $ in these limiting cases.

Finally, we present a brief discussion of thermoelectric efficiency. First, it follows from \cref{eq:reduced_efficiency} that the efficiency of heat to electricity conversion in electrolyte solutions is measured by the figure of merit $ ZT_{\mathrm{ave}} $, which is the same as that in semiconductors. The larger the figure of merit is, the higher the efficiency can reach. Thus, we only need to consider the value of $ ZT_{\mathrm{ave}} $. For an aqueous~\ce{NaCl} of 0.1~M with $ \bar{\kappa}_0\approx 1 $ (thus $ a\sim 1 $ nm) and $ \bar{\zeta}= -5 $, one estimates that $ ZT_{\mathrm{ave}} \simeq 0.53\times 10^{-3} $, which is of the same order as those obtained in non-aqueous electrolytes~\cite{Bonetti2011} although in this case the Seebeck coefficient of aqueous~\ce{NaCl} is lower than a tenth of those of non-aqueous electrolytes. It is the surface charge that enhances the effective electrical conductivity of electrolyte solution and thus improves the figure of merit. Therefore, in addition to improving the Seebeck coefficient, increasing the surface charge density also can enhance the figure of merit by increasing the effective electrical conductivity. In addition, replacing simple electrolytes with polymer electrolytes can lead to an increase in the Seebeck coefficient by tens of times~\cite{Li2019}, which potentially enhances the figure of merit.

\subsection{Analysis of fluid flow}\label{sec:flow_behavior}
\begin{figure}[!htb]
	\centering{\includegraphics[width=1\linewidth]{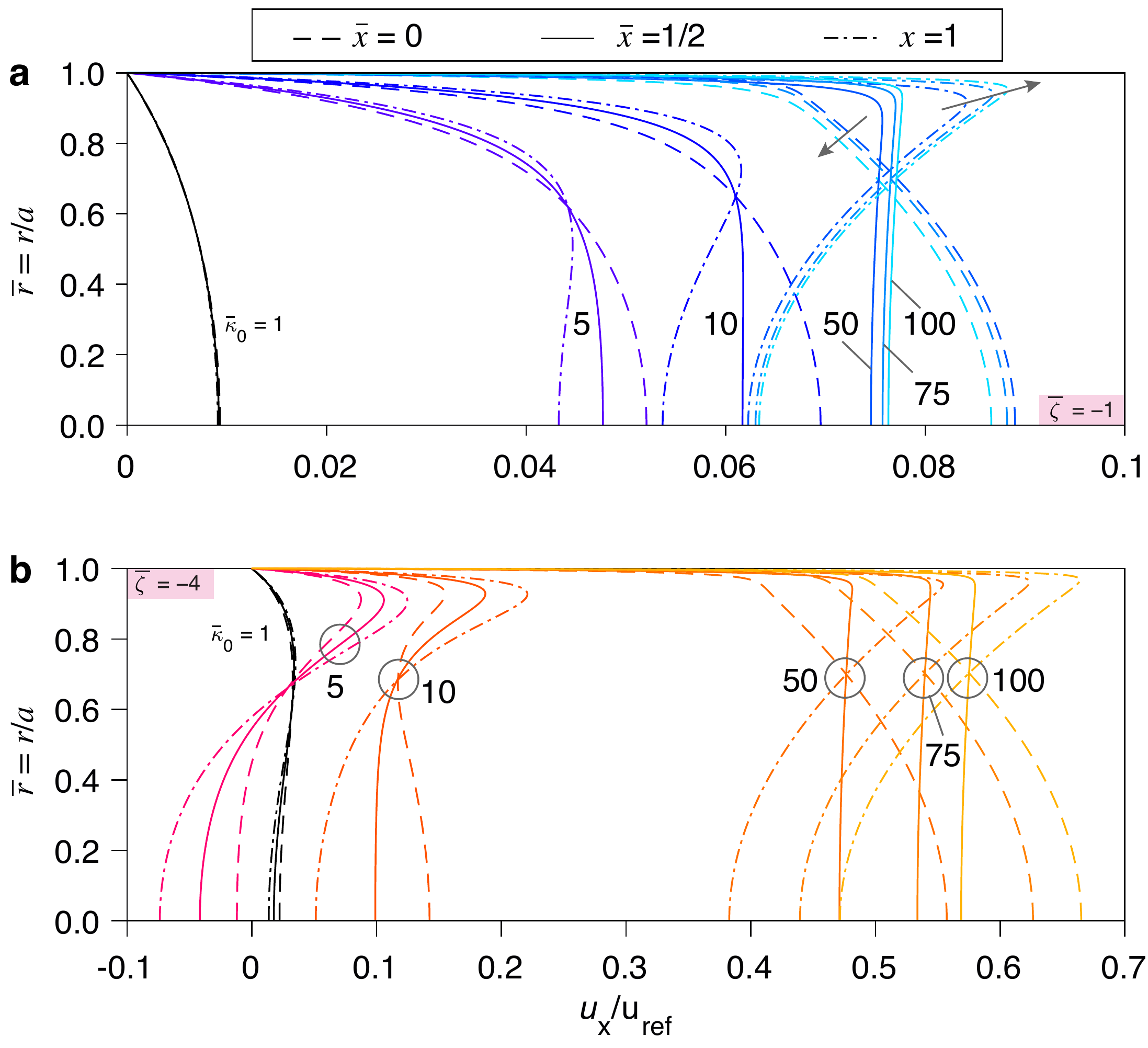}}
	\caption{Axial velocity profile $u_x$ (normalized by $u_{\mathrm{ref}}$) in \ce{NaCl} solutions at cold end (dashed lines), mid-section (solid lines) and hot end (dash-dotted lines) for (a) $\bar{\zeta}=-1$ and (b) $\bar{\zeta}=-4$ with varying $ \bar{\kappa}_0 $ (1, 5, 10, 50, 75, 100). The ambient temperature and temperature difference are set to $ T_{\infty}/T_{\mathrm{ref}}=1 $ and $ \Delta T/T_{\mathrm{ref}}=0.0671 $, respectively.}
	\label{fig:v_tot}
\end{figure}

The osmotic pressure and electric field induced by the temperature gradient can drive a flow of an electrolyte solution in a charged capillary. \cref{fig:v_tot} shows the axial velocity profiles of this flow at $ \bar{x}=0 $ (cold end), $ 1/2 $ (mid-section) and $ 1 $ (hot end) for $ \bar{\zeta}=-1 $ (a) and $ \bar{\zeta}=-4 $ (b). Evidently, for $ \bar{\kappa}_0=1 $, the velocity profiles weakly vary along the axial direction of the nanocapillary. With $ \bar{\kappa}_0 $ increasing, the variation of the velocity profiles with axial location becomes more pronounced. As $ \bar{\kappa}_0 \geq 50 $, one can clearly observe that the velocity profiles shift from convex shape at  $ \bar{x}=0 $ to concave shape at  $ \bar{x}=1 $. This behavior can be explained as follows. On the one hand, an elevation of temperature can lead to a decrease of viscosity, which can enhance the flow in the region close to the wall. On the other hand, for the \ce{NaCl} solution ($ S_{\mathrm{T}}>-\gamma-1/{T_{\mathrm{ref}}} $) an elevation of temperature would lead to a smaller $ \kappa $ value \cite{dietzel2017} and thus a thicker EDL, which can weaken the flow near the wall. Since the first effect is dominant, the velocity near the wall is larger at elevated temperature, while the velocity in the center region of the capillary decreases to keep the volumetric flow rate constant.

For low $\zeta$ potentials (e.g., $\bar{\zeta}=-1$), the velocity profiles at the capillary mid-section are analogous to those of the conventional EOF under isothermal conditions to some extent, i.e., exhibiting a parabolic structure for small values of $\bar{\kappa}_0 $ and a plug-like structure for large values of $\bar{\kappa}_0 $ (\cref{fig:v_tot}a). Also, in the center of the nanocapillary with increasing $\bar{\kappa}_0$, the fluid velocity magnitude gradually increases, and eventually saturates to a constant as $\bar{\kappa}_0 \tti$. For high $\zeta$ potentials (e.g., $\bar{\zeta}=-4$), the velocity profiles are quite different from the low $ \zeta $ potential, as is shown in \cref{fig:v_tot}b. For very small ($ \bar{\kappa}_0 \leq 1 $) and very large ($ \bar{\kappa}_0 \geq 10 $) values of $ \bar{\kappa}_0 $, the velocity is always directed to the $+x$ direction, namely directed to the hot end. However, for moderate values of $ \bar{\kappa}_0 $, a reversal in the velocity direction is present in the center of the nanocapillary (i.e., backflow), while in the region close to the nanocapillary wall the velocity is still directed toward the hot end. This backflow is found to be a non-monotonic function of $\bar{\kappa}_0$; it achieves the maximum extent at a certain $\bar{\kappa}_0$ and decreases as $\bar{\kappa}_0$ deviates from this value. Because of such backflow, there would be a presence of a cylindrical stationary surface inside the nanocapillary on which the fluid velocity is zero. Therefore, on this surface the ion motion is in nature caused by the thermophoresis alone. Potentially, this feature can be used for the investigation of the thermophoretic ion motion. In the limit of $\bar{\kappa}_0 \tti$, the flow velocity saturates to a constant as in the case of low $ \zeta $ potentials.

\begin{figure*}[!htb]
	\centering{\includegraphics[width=0.75\linewidth]{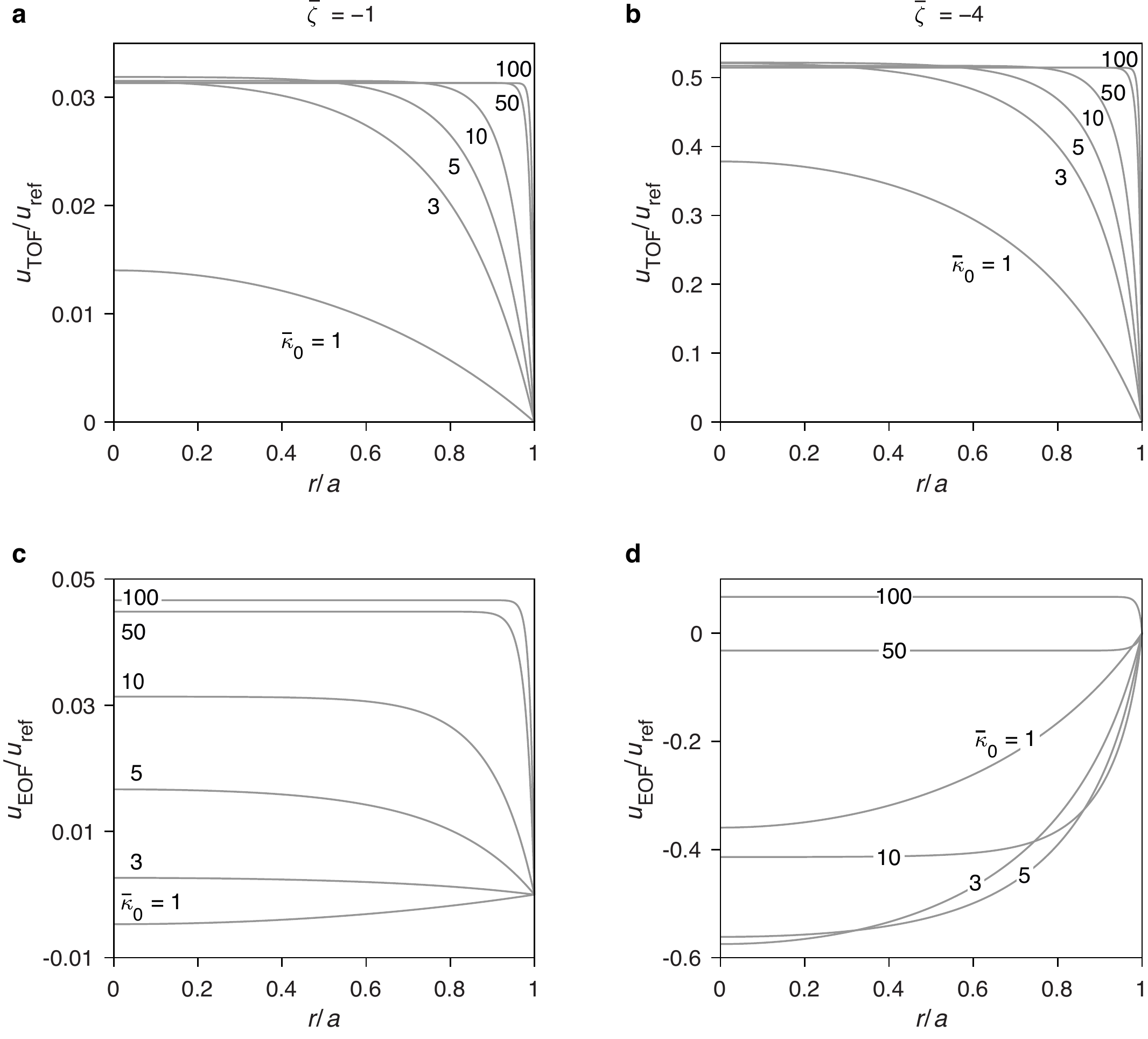}}
	\caption{Axial velocity profiles of the TOF component (a, b) and the EOF component (c, d) of fluid flow in \cref{fig:v_tot} at the capillary mid-section. The $\zeta$ potential is set to $\bar{\zeta}=-1$ in (a, c) and $\bar{\zeta}=-4$ in (b, d).}
	\label{fig:v_tof_eof}
\end{figure*}

The temperature-gradient-driven electrokinetic flow can be decomposed as a thermoelectric-field-induced EOF given by the third term of \cref{eq:u_x} and a TOF given by sum of the second and last terms of \cref{eq:u_x}. The TOF is in nature driven by the excess osmotic pressure (termed thermo-chemioosmotic flow \cite{Maheedhara2018}) and the dielectric body force (expressed by the term proportional to $ \gamma $)~\cite{wurger2010}. To have a more detailed understanding of the two flow components, \cref{fig:v_tof_eof} presents the TOF (a, b) and EOF (c, d) velocity components of the total flow at the capillary mid-section (where $ \p p_0 /\p x $ would be negligibly small) in \cref{fig:v_tot}. Clearly, the TOF velocity profiles resemble the conventional EOF under isothermal conditions (\cref{fig:v_tof_eof}a, b). By contrast, the behaviors of the EOF are more complicated. For $\bar{\zeta}=-1$, the velocity profiles of the EOF are analogous to those of the conventional EOF to some extent, evolving from a parabolic to a plug-like structure as $\bar{\kappa}_0 $ increases (\cref{fig:v_tof_eof}c). Additionally, the magnitude of the ``plug-like'' velocity in the center of capillary gradually increases with $\bar{\kappa}_0 $ ($ \gtrsim 3 $), and eventually reaches a plateaued value as $\bar{\kappa}_0 \tti $. This is because in this case the thermoelectric field that drives the EOF becomes stronger as $\bar{\kappa}_0$ increases and finally reaches the bulk Soret thermoelectric field as $\bar{\kappa}_0\tti$ (\cref{sec:te_effect}). For $\bar{\zeta}=-4$ with $\bar{\kappa}_0\leq \bar{\kappa}_{0,cr} $ ($ \simeq 60.9 $), we find that the EOF drives the solvent to the cold end (namely, the velocity is negative as shown in \cref{fig:v_tof_eof}d). As $\bar{\kappa}_0$ increases within $ 1 \leq \kappa \leq \bar{\kappa}_{0,cr} $, it is seen that the velocity magnitude of this EOF initially increases and then decreases. For $\bar{\kappa}_0 \geq \bar{\kappa}_{0,cr} $, the velocity even switches direction. This is because increasing the value of $\bar{\kappa}_0 $ on the one hand has the tendency to enhance the EOF, while on the other hand has the tendency to reduce the thermoelectric field and then the EOF as long as $ \bar{\kappa}_0<\bar{\kappa}_{0,cr} $ (\cref{fig:Seebeck}a). Clearly, increasing $\bar{\kappa}_0 $ has two contradicting effects on the EOF. Thus, the competition between these two contradicting effects results in an optimal value of $\bar{\kappa}_0 $ such that the EOF is strongest.

After having a clear understanding of the characteristics of individual flow components, namely the TOF and the EOF, one then can easily understand the characteristics of total flow in \cref{fig:v_tot}. For low $\zeta$ potentials or for high $\zeta$ potentials with $ \bar{\kappa}_0>\bar{\kappa}_{0,cr} $, the directions of the two flow components are consistent, jointly driving the fluid to the hot end.
For high $\zeta$ potentials with $ \bar{\kappa}_0<\bar{\kappa}_{0,cr} $, the TOF velocity and EOF velocity are in opposite directions, and the characteristics of overall velocity is then determined by the interaction between these two opposing flow components.

\begin{figure}[!htb]
	\centering{\includegraphics[width=0.75\linewidth]{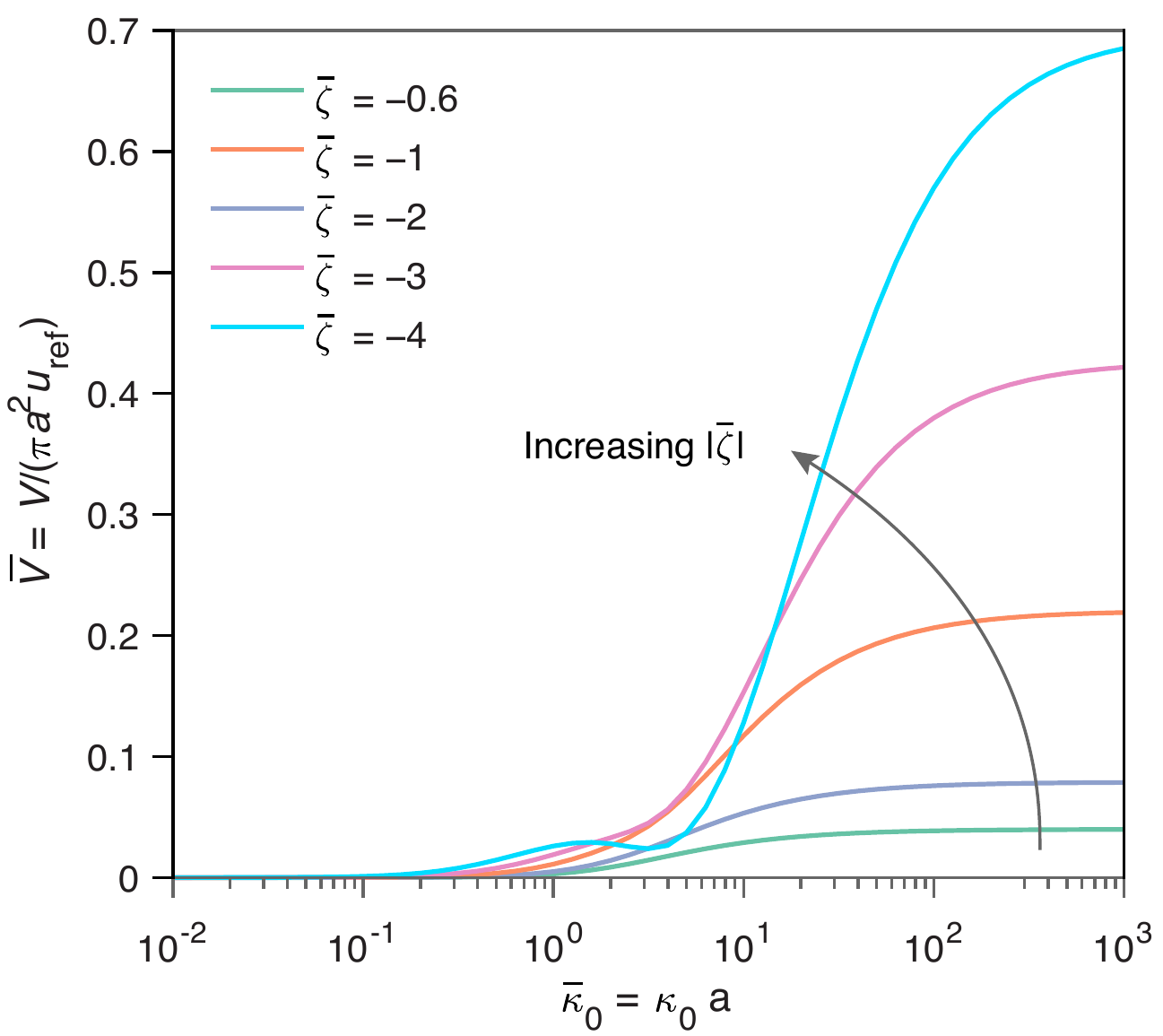}}
	\caption{Dimensionless average velocity $\bar{V}$ as a function of $\bar{\kappa}_0 $ for the aqueous solution of NaCl with $ \bar{\zeta}=-0.6,\;-1,\;-2,\;-3 $ and $ -4 $. The temperature conditions are the same as \cref{fig:v_tot}.}
	\label{fig:average_velocity}
\end{figure}

We next analyze the average velocity $\bar{V}$ in the \ce{NaCl} solution. Clearly, for $ |\bar{\zeta}| $ lower than $ 3 $ with increasing $ \bar{\kappa}_0 $, $\bar{V}$ increases slowly when $ \bar{\kappa}_0 \lesssim 1 $, while increases sharply within $ 1 \lesssim \bar{\kappa}_0 \lesssim \mathcal{O}(10) $ and eventually saturates to \cref{eq:average_velocity_inf} as $\bar{\kappa}_0 a\tti$ (\cref{fig:average_velocity}). By contrast, for $ \bar{\zeta}=-4 $ the average velocity is no longer a monotonic increasing function of $ \bar{\kappa}_0 $. Instead, with the increase of $ \bar{\kappa}_0 $, $\bar{V}$ increases first, then decreases and increases again up a plateau value as $ \bar{\kappa}_0\tti $. The fluctuation of $\bar{V}$ within a certain $ \bar{\kappa}_0 $ range originates from the competition between the EOF and the TOF. This competition also results in the fact that $\bar{V}$ is a non-monotonic function of $ |\bar{\zeta}| $.

\begin{figure}[!htb]
	\centering
	\includegraphics[width=1\linewidth]{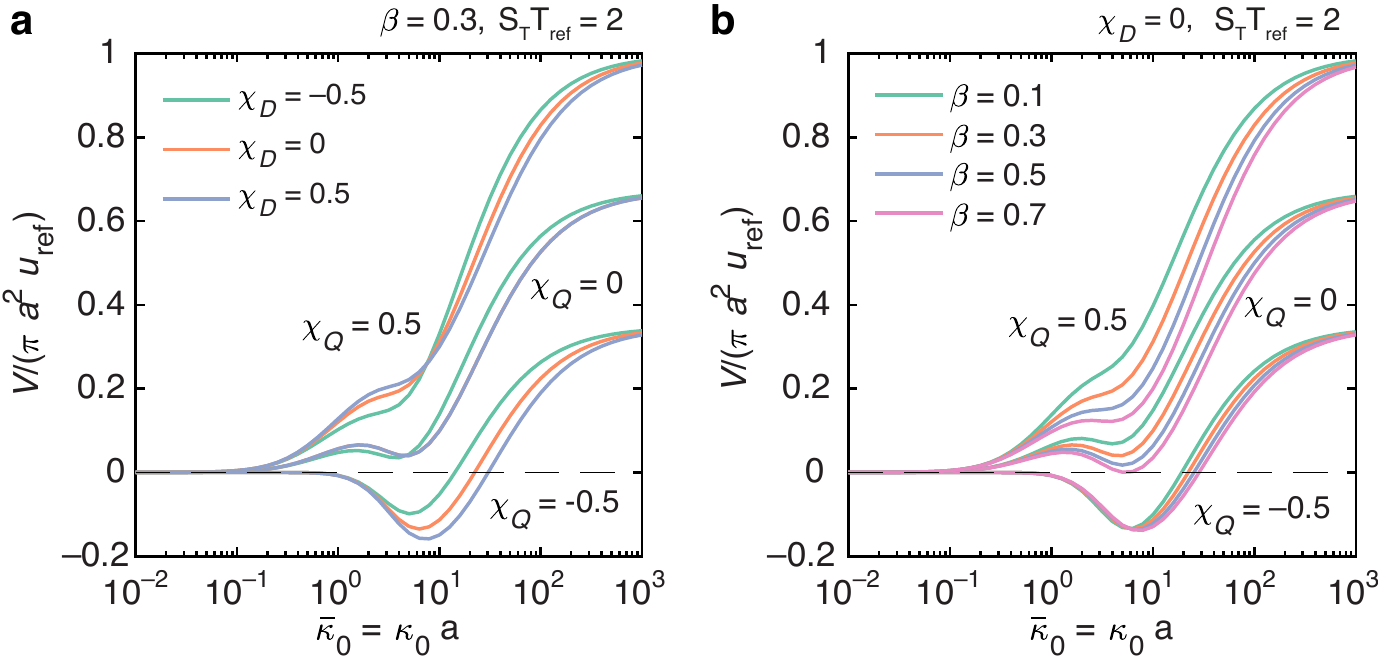}
	\caption{Dimensionless average velocity $ \bar{V} $ as a function of dimensionless Debye parameter $ \bar{\kappa}_0 $ for (a) varying $ \chi_{D} $ and $ \chi_{Q} $ with $ \beta=0.3 $, $ S{T_{\mathrm{ref}}}=2 $ and (b) varying $ \beta $ and $ \chi_{Q} $ with $ \chi_{D}=0 $, $ S{T_{\mathrm{ref}}}=2 $. The $ \zeta $ potential is set to $ \bar{\zeta}=-4 $. The temperature conditions are the same as \cref{fig:v_tot}.}
	\label{fig:parametric_v}
\end{figure}

We then analyze the effects of three parameters ($ \chi_{D} $, $ \beta $ and $ \chi_{Q} $) on the average velocity $ \bar{V} $. To this end, \cref{fig:parametric_v}a and b shows $ \bar{V} $ as a function of $ \bar{\kappa}_0 $ for varying $ \chi_{D} $ and varying $ \beta $, respectively, with $ \chi_{Q} =(-0.5,\;0,\; 0.5) $, $ \bar{\zeta}=-4 $ and $ ST_{\mathrm{ref}}=2 $. Clearly, neither $ \chi_{D} $ nor $ \beta $ has an effect on the average velocity as $ \bar{\kappa}_0\ll 1 $ or $ \bar{\kappa}_0\gg 1 $ (\cref{fig:parametric_v}). For moderate $ \bar{\kappa}_0 $ values, both $ \chi_{D} $ and $ \beta $ have substantial influence on the average velocity. It is apparent from \cref{fig:parametric_v}a that for $ \chi_{Q}=0 $ there exists a transition point for $ \bar{\kappa}_0 $ (ca. 5) such that the average velocity is almost independent of $ \chi_{D} $. If $ \bar{\kappa}_0 $ is below (above) such transition point, the average velocity increases (decreases) with increasing $ \chi_{D} $. For $ \chi_{Q}=0.5 $ the characteristics of the average velocity are similar to the case of $ \chi_{Q}=0 $, but the corresponding transition point for $ \bar{\kappa}_0 $ shifts to ca. 8. By contrast, for $ \chi_{Q}=-0.5 $ the transition point for $ \bar{\kappa}_0 $ vanishes. In this case, a backflow is observed at certain range of $ \bar{\kappa}_0 $ ($ \simeq 1 $--$ \mathcal{O}(10) $). Such backflow is enhanced by increasing the value of $ \chi_{D} $. \cref{fig:parametric_v}b shows that for $ \chi_{Q}=0 $, $ \bar{V} $ decreases with increasing $ \beta $ and such trend becomes more (less) pronounced as $ \chi_{Q} $ varies from 0 to 0.5 ($ -0.5 $). 
It is noted that the aforementioned characteristics of the average velocity are quite similar to the thermoelectric field in \cref{sec:te_effect}. This is understandable since the three parameters ($ \chi_{D} $, $ \beta $ and $ \chi_{Q} $) affect the average velocity by altering the thermoelectric field. 

In contrast to the above electrolyte-related parameters, $ ST_{\mathrm{ref}} $ has a substantial effect on both the TOF and EOF components of the average velocity, which can be deduced straightforwardly from \cref{average_velocity} and is not discussed here. 

\begin{figure*}[!htb]
	\centering
	\includegraphics[width=0.75\linewidth]{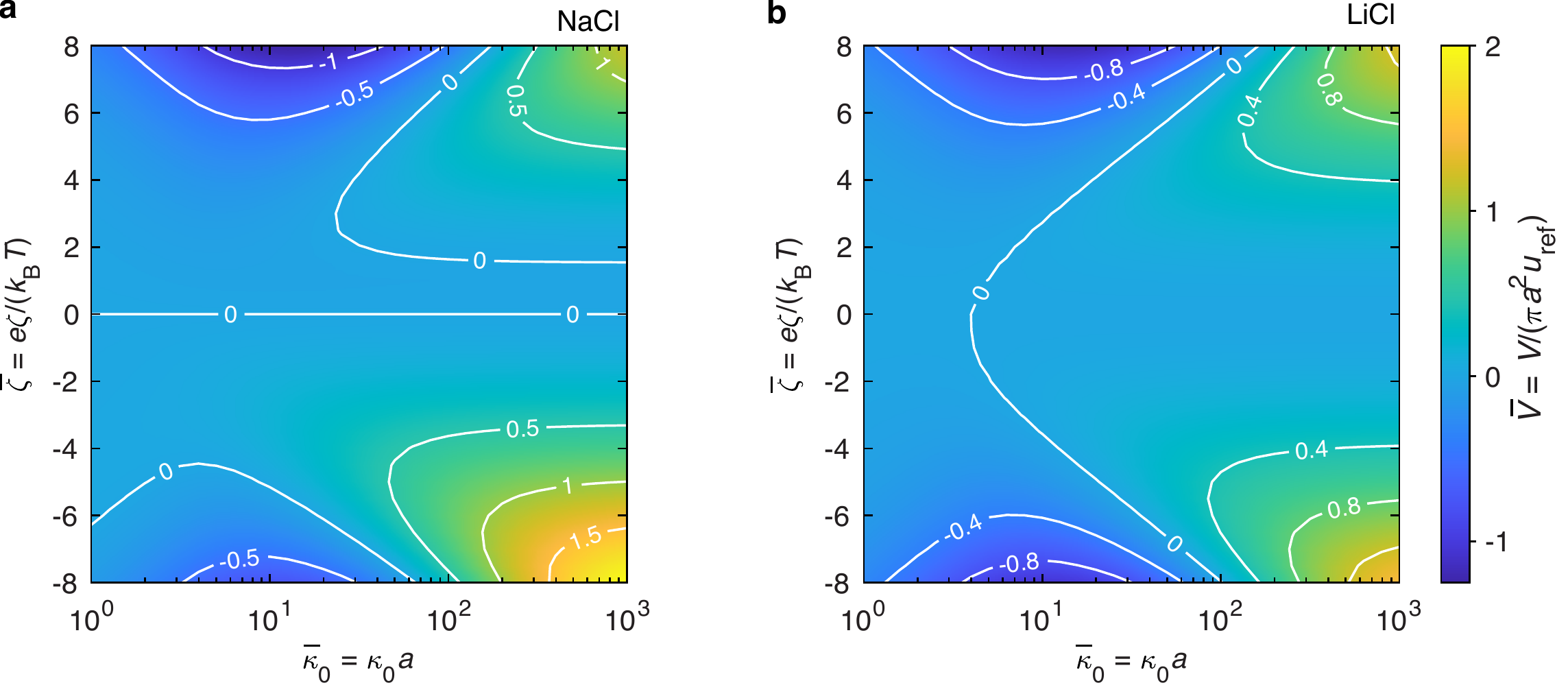}
	\caption{Contour plots of average velocity $ \bar{V} $ for (\textit{a}) NaCl and (\textit{b}) LiCl in a $ \bar{\kappa}_0 $--$ {\zeta}^{\ast} $ domain. The temperature conditions are the same as \cref{fig:v_tot}.}
	\label{contour2}
\end{figure*}

Finally, to have a more comprehensive view of the effects of $ \bar{\kappa}_0 $ and $ \bar{\zeta} $ on the average velocity for various practical electrolytes with electrolyte-specific parameters, in \cref{contour2} we display contour plots of $ \bar{V} $ in a $ \bar{\kappa}_0 $--$ \bar{\zeta} $ domain for (a)~\ce{NaCl} and (b)~\ce{LiCl}. Here, we only focus on the cases in which $ 1 \leq  \bar{\kappa}_0 \leq 1000 $ and $ |\bar{\zeta}|\leq 8 $. It is apparent from \cref{contour2}a that $\bar{\zeta}$ and $\bar{\kappa}_0 $ have complicated effects on the average velocity of~\ce{NaCl} solution. For $ -4.5 \lesssim \bar{\zeta} \lesssim 1.5 $, the sign of $ \bar{V} $ is opposite to that of $ \bar{\zeta} $. For $ \bar{\zeta} \gtrsim 1.5 $, $ \bar{V} $ is negative below certain values for $ \bar{\kappa}_0 $ ($ \gtrsim \mathcal{O}(10^1) $). For $\bar{\zeta} \lesssim -4.5 $, $ \bar{V}<0 $ is present within a certain range of $ \bar{\kappa}_0  $ (approximately $ \mathcal{O}(1) $--100). The higher the $ |\bar{\zeta}| $ value is, the broader such range of $ \bar{\kappa}_0 $ becomes. In comparison, for a~\ce{LiCl} solution, the contour plot in \cref{contour2}b is more regular since it is nearly symmetric with the line of $ \bar{\zeta}=0 $. For nonzero values of $ \bar{\zeta} $, $ \bar{V}<0 $ ($ \bar{V}>0 $) occurs below (above) a certain value of $ \bar{\kappa}_0 $. The features in \cref{contour2} can be understood by analyzing the intricate interplay between two velocity components, namely the TOF and the EOF. The difference between \cref{contour2}a and b is found to be mainly caused by the difference in the intrinsic Soret coefficients between~\ce{Na+} and~\ce{Li+}.

\section{Conclusions}\label{sec:conclusions}

The present work preforms a systematic analysis of the thermoelectric potential and temperature-gradient-driven electrokinetic flow of electrolyte solutions in charged capillaries. The analysis is based on a semi-analytical model developed by solving the non-isothermal PNPNS equations with the lubrication theory. As a generalized theory, the model agrees with the classic isothermal capillary pore model when the temperature gradient vanishes, and can exactly recover widely-accepted theories in limiting cases. In contrast to previous works, this work builds a connection between the thermoelectric potential as well as the temperature-gradient-driven electrokinetic flow and membrane properties (effective pore radius and length, surface potential or charge density) to have deep insights into the electrolyte and water transport in charged pores driven by the temperature gradient. The major conclusions are summarized as follows.

This paper clarifies the interplay and relative importance of three thermoelectric mechanisms, i.e., (1) the ion convection due to the TOF, (2) the ion selective diffusion caused by the temperature dependence of ion electrophoretic mobility, and (3) the ion thermodiffusion due to the difference in the Soret coefficients between cation and anion, under  different combinations of $ \zeta $ potential and dimensionless Debye parameter $ \bar{\kappa}_0 $. The Seebeck coefficient has a electrolyte-dependence, which is reflected by the effects of four variables, i.e., the normalized difference in the diffusion ($ \chi_{D} $) and Soret coefficients ($ \chi_{Q} $) between cation and anion, the intrinsic P{\'e}clet number $ \beta $ and the reduced Soret coefficient of cation and anion $ S_{\mathrm{T}} T_{\mathrm{ref}} $. Besides, we proposed synergy conditions for three thermoelectric mechanisms to fully cooperate, i.e., $ \chi_{Q}\zeta >0 $ and $ \chi_{Q}\zeta <0 $ for thermophobic and thermophilic electrolytes respectively, and for the thermophilic electrolytes an extra constraint is required. Additionally, our results indicate that the Seebeck coefficient is affected by the ambient temperature. Next, we show that the temperature-gradient-driven electrokinetic flow is a nearly unidirectional flow whose axial velocity profiles vary with the axial location. Essentially, this flow is a consequence of the interplay of a TOF and an EOF. These two flow components have intricate dependence on $ \zeta $ potential and $ \bar{\kappa}_0 $. Similar to the Seebeck coefficient, the average velocity is also affected by the electrolytes through four electrolyte-related variables, $ \chi_{D} $, $ \chi_{Q} $, $ \beta $ and $ S T_\mathrm{ref} $. The first three variables have almost no effect on the TOF component of the average velocity, while affect appreciably the EOF component of the average velocity by adjusting the magnitude and direction of the thermoelectric field. By contrast, $ S_{\mathrm{T}} T_\mathrm{ref} $ affects both components of the average velocity substantially.

The findings in this work are beneficial for understanding the underlying physical mechanisms of the fluid flow and electric field induced along charged capillaries by the temperature gradient. The model developed in this study is also helpful for designing low-grade waste heat recovery devices and thermoosmotic pumps. In the future, one can include other effects in the current model, such as temperature dependence of $ \zeta $ potential \cite{revil1999}, temperature dependencies of Soret coefficients \cite{iacopini2003}, slip on shear plane \cite{eijkel2007}, finite size of ions \cite{borukhov1997}, ion correlation \cite{bazant2011} and activity coefficients. These effects are of high importance under some extreme conditions, such as very large temperature gradient, extreme confinement, high salt concentration and high surface charge density etc. {In addition, the model can be further modified to take into account the variable temperature gradient along the membrane surface \cite{Sandbakk2013} since the membrane has a thermal conductivity the same order of magnitude as aqueous solutions.}

\section*{Acknowledgments}
	We acknowledge the support received from the National Key Research and Development Program of China (No. 2017YFB0603500), National Natural Science Foundation of China (No. 51721004, No. 51976157).
\section*{Conflict of interest}
The authors declare that there is no conflict of interest.
\appendix

\section{Order-of-Magnitude analysis}\label{appA}
In two-dimensional axisymmetric geometry, the viscous dissipation is given as $ \dot{\Phi}_{\mu}=2\mu[(\p_r u_r)^2+(u_r/r)^2+(\p_x u_x)^2+(\p_r u_x+\p_x u_r)^2/2] $, which can be approximated by $ \dot{\Phi}_{\mu}\approx \mu (\p_r u_x)^2  $ under the conditions of $ a/l\ll 1 $ (it is reasonable since in this work we consider capillaries with radius of a few to dozens of nanometers and length of a few microns), where all variables and coefficients are defined in the main text. Then, the relative importance of the viscous dissipation and the conductive heat transfer is characterized by the Brinkman number $ Br=\mu_{\mathrm{ref}} u_{\mathrm{ref}}^2/(k_{\mathrm{ref}}T_{\mathrm{ref}}) $~\cite{ghonge2013}.

The relative magnitude of the convective to the conductive heat transfer is measured by the thermal P{\'e}clet number $ Pe_T= u_{\mathrm{ref}} a/a_T $ with $ a_T $ being the thermal diffusivity of the electrolyte solution. Similarly, the relative importance of the ion convection to the ion diffusion is characterized by the ionic P{\'e}clet number $ Pe_i=u_{\mathrm{ref}} a/D_{\mathrm{ref}}$.

With $ T_{\mathrm{ref}}=298 $ K, $ \epsilon_{\mathrm{ref}}\sim 6.95\times 10^{-10}\;$F m$^{-1} $, $ \mu_{\mathrm{ref}}\sim 8.9\times 10^{-4}\;$Pa s$ $, $ a_T\sim 1.47\times 10^{-7}\;$m$^2\;$s$^{-1} $, $ D_{\mathrm{ref}}\sim 10^{-9}\;$m$^2\;$s$^{-1} $, $ l\sim \mu $m and $ u_{\mathrm{ref}}\sim (\epsilon_{\mathrm{ref}}/\mu_{\mathrm{ref}} l)(k_{\mathrm{B}}T_{\mathrm{ref}}/e)^2 $, one estimates that $ Br\sim O(10^{-9})\ll 1 $, $ Pe_T\sim O(10^{-3}a/l)\ll 1 $ and $ Pe\sim O(0.1a/l)\ll1 $. Thus, within the scope considered in this study both the viscous dissipation and the convective heat transfer are negligible, and the contribution to the ion distribution in the EDL due to the ion convection also can be neglected.

Next, to show that the axial gradient of the viscosity does not contribute to the semi-analytical solution, we expand the Navier-Stokes equation as
\begin{widetext}
\begin{align}
-\frac{\p p}{\p x}+\frac{1}{r}\frac{\p}{\p r}\left( r \mu \frac{\p u_x}{\p r} \right) +\mu \frac{\p^2 u_x}{\p x^2}+2\frac{\p u_x}{\p x} \frac{\p \mu}{\p x}-\rho_e \frac{\p \phi}{\p x} -\frac{\gamma \epsilon}{2} \left[  \left( \frac{\p \phi}{\p x}\right)^2+ \left( \frac{\p \phi}{\p r}\right)^2 \right] \frac{\p T}{\p x}=0&\\-\frac{\p p}{\p r} +\mu \left[\frac{\p}{\p r} \left(  \frac{1}{r} \frac{\p (r u_r)}{\p r} \right)  +\frac{\p^2 u_r}{\p x^2}\right] + \left( \frac{\p u_x}{\p r}+\frac{\p u_r}{\p x}\right)  \frac{\p\mu}{\p x} -\rho_e \frac{\p \phi}{\p r}=0&
\end{align}	
\end{widetext}
in which the terms multiplied with $ \p_r \mathscr{F} $ have been omitted, where $ \mathscr{F}\equiv(T,\;\epsilon,\;\mu) $.

Under the assumption of $ a/l\ll 1 $, one can estimate that $ \mu\p_x^2 u_x \sim  \p_x u_x \p_x \mu \ll r^{-1}\p_r(r\mu\p_r u_x) \sim \p_x p \sim \rho_e\p_x\phi \sim \gamma\epsilon (\p_r\phi)^2\p_x T $ and $ \mu\p_x^2 u_r\sim \p_xu_r \p_x \mu \ll \mu \p_r(r^{-1}\p_r(r u_r)) \sim \p_r u_x \p_x \mu \ll \p_r p\sim \rho_e \p_r \phi $. Therefore, the terms multiplied with $ \p_x\mu $ can be reasonably neglected and simultaneously \cref{eq:v_x,eq:v_r} are derived.

\section{Physical properties and their temperature dependences}\label{appB}

The dynamic viscosity, the dielectric permittivity and the thermal conductivity of the electrolyte solutions are taken as those of water and their temperature dependencies can be fitted with tabulated data \cite{haynes2014} as $ \mu(T)=\mu_{\mathrm{ref}}\exp[-(b_1(T-{T_{\mathrm{ref}}})+b_2(T-{T_{\mathrm{ref}}})^2)/(T-b_3)] $, $ \epsilon(T)=\epsilon_{\mathrm{ref}}\exp[\gamma(T-{T_{\mathrm{ref}}})] $ and $ k(T)=k_0\exp[-(T-c_1)^2 /c_2^2] $  ($ 273.16 \;$K$ \leq T \leq 372.756 \;$K$ $), where $ \mu_{\mathrm{ref}}=8.9\times 10^{-4}\;$Pa s$ $, $ \epsilon_{\mathrm{ref}}=6.95\times 10^{-10}\;$ F~m$^{-1} $, $ b_1=2.634 $, $ b_2=0.003744\;$K$^{-1} $, $ b_3=183\;$K$ $, $ \gamma=-1/216.9\;$K$^{-1} $, $ k_0=0.6788\;$W m$^{-1}\;$K$^{-1} $, $ c_1=379.9\;$K$ $ and $ c_2=244.7\;$K$ $. Additionally, we relate the ion diffusivity and temperature with the Stoke-Einstein relationship by assuming the hydrostatic radii of ions are independent of temperature \cite{caldwell1981,dietzel2017}, i.e., $ \mu D_{\pm}/T=\,$const$ $.
Furthermore, Diffusion and Soret coefficients of three aqueous solutions of alkali chloride (LiCl, NaCl and KCl) are summarized in Table \ref{tab:parameter}.
\renewcommand{\thetable}{B\arabic{table}}
\begin{table}[!htb]
\centering
\caption{\label{tab:parameter}Diffusion coefficients $D_i$ and molar heat of transports $Q_i^{\ast}$ for at infinite dilution, and the corresponding values for salts. All values are determined at $25\;^{\circ}$C ($ \sim $298 K). The values $D_i$ are taken from ref. \cite{haynes2014}, while $Q_i$ from ref. \cite{agar1989}. In addition, $\mathscr{S}_{\mathrm{T}}=S_{\mathrm{T}}T_{\mathrm{ref}}=(Q_++Q_-)/(2N_{\mathrm{A}} k_{\mathrm{B}} T_{\mathrm{ref}})$ is the average reduced Soret coefficient of the electrolyte with $N_{\mathrm{A}}
		$ being the Avogadro constant. These values are in units of ($a$) $10^{-9}\;$m$^2\;$s$^{-1} $ and ($b$) $\;$kJ\,mol$^{-1}$.}
\begin{ruledtabular}
\begin{tabular}{lcclcccc}
Ion      & $D_i$$^{(a)}$  & $Q_i$$^{(b)}$   & Salt & $ \beta $  & $\chi_{D}$  & $ \mathscr{S}_{\mathrm{T}}$  &  $\chi_{Q} $   \\
\hline
$\ce{Li+}$  & 1.029  & 0.53           & \ce{LiCl}      &    0.3362             &  -0.328 &  0.214                   &  0~~~             \\
$\ce{Na+}$  & 1.334  & 3.46          & \ce{NaCl}     &      0.3058            &  -0.207 &  0.805                   &  0.735             \\
$\ce{K+} $  & 1.957  & 2.59          & \ce{KCl}       &    0.2580             &  -0.019 &  0.625                   &  0.643             \\
$\ce{Cl-}$  & 2.032  & 0.53           &      &                  &                      &                                 &                   \\
\end{tabular}
\end{ruledtabular}

\end{table}

\section{Validation}\label{appC}
\renewcommand{\thefigure}{C\arabic{figure}}
\setcounter{figure}{0}
It is clear that in the absence of temperature gradient, \cref{average_velocity,eq:current} can fully recover the classic isothermal capillary pore model (or space charge model) \cite{Gross1968}. Alternatively, \cref{average_velocity,eq:current} can also be rewritten in terms of $ (-\p_x p_t,\; -k_{\mathrm{B}}T \p_x \ln n,\; \p_x \varphi,\; \p_x \ln T )^{\mathrm{T}} $ instead of $ (-\p_x p_0,\; -k_{\mathrm{B}}T \p_x \ln n,\; \p_x \varphi,\; \p_x \ln T)^{\mathrm{T}} $, where $ p_t=p_0-2 n k_{\mathrm{B}}T $ is the virtual total pressure (also termed solvent pressure). The reformed expressions for $ V $ and $ I $ are in agreement with another version of the capillary pore model \cite{peters2016} when $ \nabla T=0 $.

For an uncharged capillary, the thermoelectric potential from \cref{eq:varphi_diff} reduces to bulk Soret voltage $ \Delta \varphi|_{\zeta\ttz}=-S_{e}^{\infty} \Delta T $ with $ S_{e}^{\infty}=\chi_{Q} k_{\mathrm{B}} S T_{\mathrm{ave}} /e $. In the limit of $ \bar{\kappa}_0 \rightarrow \infty $, we have $ L_2\to-\bar{\zeta}\epsilon/\epsilon_0 $, $ L_5\to\chi_{D}/2 $, $ L_7\to1/2 $, $ L_9\to\chi_{Q}/2 $ and $ L_i\to 0 $ for other $ i $, then $ C_0=C_1=C_3=0 $, $ C_2=1 $ and therefore $ \Delta \varphi|_{\bar{\kappa}_0\tti}=-S_{e}^{\infty} \Delta T $. In the limit of $ \bar{\kappa}_0\ttz $, one has $ C_0=C_1=0 $, $ C_2=1 $ and $ C_3=\Psi_s $, thus $ \Delta \varphi|_{\bar{\kappa}_0\ttz}=-S_{e}^{\infty} \Delta T -\zeta\ln(1+\Delta T/T_{\infty}) $. All these results are consistent with those of slit channels \cite{dietzel2017}, which is reasonable since in these cases the geometries do not play a role.

In the limit of $\bar{\kappa}_0 \tti$, the average velocity $V$ saturates to (under the Debye-H\"uckel approximation)
\begin{widetext}
\begin{align}\label{eq:average_velocity_inf}
V_{\infty}=&\frac{1}{8 \langle\mu\rangle} \int_{T_{\infty}}^{T_{\infty}+\Delta T} \left[ \frac{\epsilon \zeta^2}{l} \left( \frac{1}{T} -\frac{\p_T \epsilon}{\epsilon} -\frac{\p_T n}{n} \right)-\frac{8\chi_{Q}\epsilon \zeta S_{\mathrm{T}}}{l} \frac{k_{\mathrm{B}} T}{ e}\right] \mathrm{d} T,
\end{align}	
\end{widetext}
which is expected to be identical to the expression for average velocity in slit channels under the same limit \cite{dietzel2017} when $ \chi_{Q}=0 $, where $ \langle\mu\rangle=\int_{0}^{l}\mu \mathrm{d} x/l $. This is reasonable since a capillary and a slit channel are effectively no difference under the limit of $\bar{\kappa}_0\tti$. Additionally, the previous studies on the ``generalized'' TOF, $ V_{\infty} $ is in general expressed as $ V_{\infty}=M_{to}\nabla T/T \approx M_{to}\nabla T/T_{\mathrm{ave}} $ with $ M_{to} $ being the thermoosmosis coefficient \cite{bregulla2016,ganti2017}. For the NaCl solution with $ \zeta\simeq -k_{\mathrm{B}}{T_{\mathrm{ref}}}/e $, we can evaluate $ M_{to} \sim 10^{-10}\;$m$^2$\,s$^{-1} $, agreeing with the experimental data reported by Bregulla~\textit{et al.}~\cite{bregulla2016} in both the sign and the order of magnitude.

\begin{figure*}[!htb]
	\centering{\includegraphics[width=0.75\linewidth]{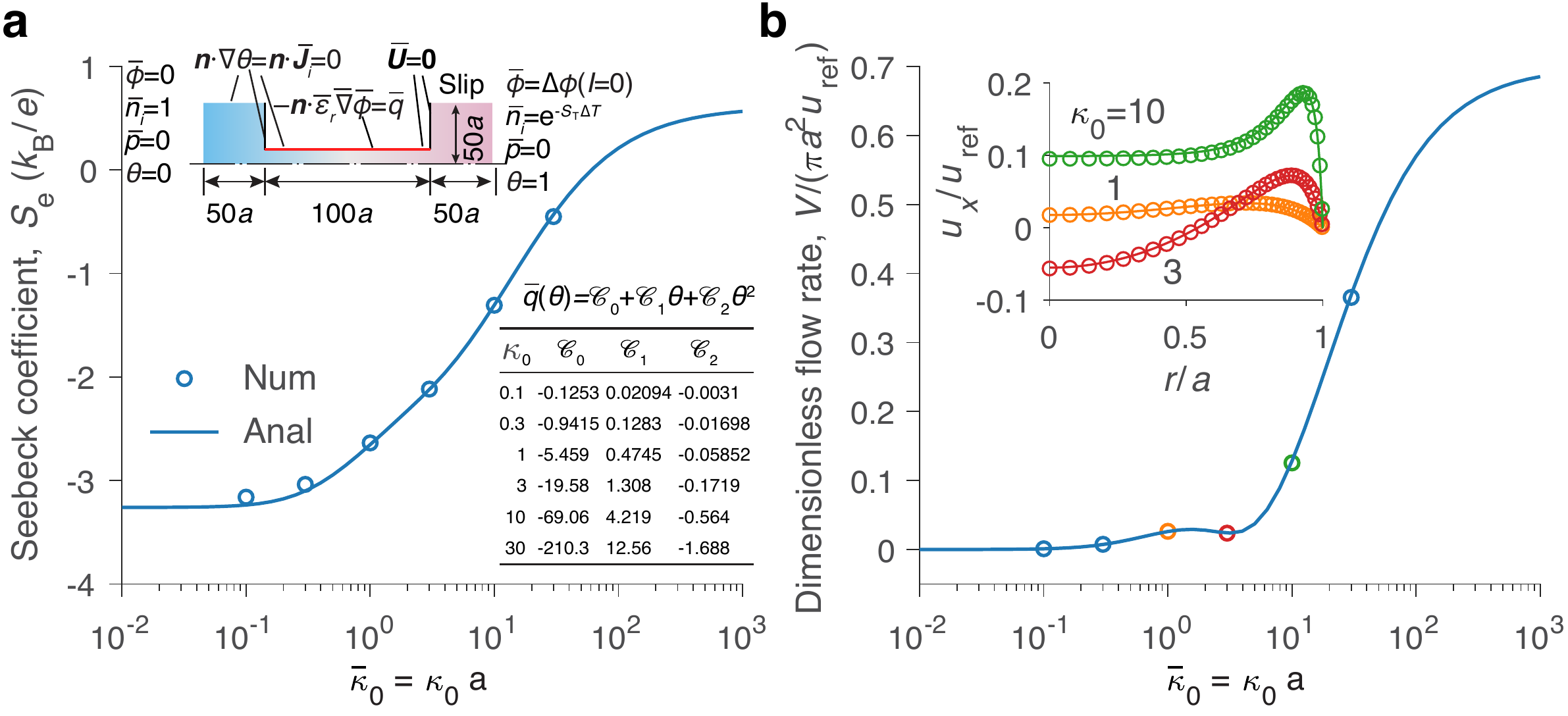}}
	\caption{Comparison between results obtained from numerical simulation (``Num'') and those calculated from the derived analytical equations (``Anal'') for $ \bar{\zeta}=-4 $ and aqueous \ce{NaCl}. The temperature conditions are the same as \cref{fig:Seebeck}. (a) Seebeck coefficient $ S_e $ versus dimensionless Debye parameter $ \bar{\kappa}_0 $. The inset shows the simulation domain and boundary conditions. All variables shown are nondimensional but share the same symbols as the dimensional variables (with bars) except $ \theta=(T-T_{\infty})/\Delta T $. (b) Flow rate $ \bar{V} $ versus dimensionless Debye parameter $ \bar{\kappa}_0 $. Inset: axial velocity profiles at $ \bar{x}=1/2 $ for $ \bar{\kappa}_0=1,\;3 $ and 10. In our numerical simulation, at the capillary wall $ \bar{q} =\bar{q} (\theta) $ (shown in the table in panel a) was employed.}
	\label{fig:validation}
\end{figure*}

Finally, we present a comparison between the numerical solutions based on \crefrange{nernst_planck_equation}{eq:nse} coupled with energy equation (detailed later in this Appendix) and the derived semianalytical solutions. Such comparison for the case of $ \bar{\zeta}=-4 $ is shown in \cref{fig:validation}. It is apparent from \cref{fig:validation} that the derived semianalytical solutions fully agree with the numerical solutions.

In the numerical simulation (COMSOL Multiphysics), the governing equations were nondimensionalized in accordance with Ref.~\cite{wood2016} except temperature (which was normalized as $ \theta=(T-T_{\infty})/\Delta T $). The temperature dependencies of physical properties introduced in Appendix~\ref{appB} were employed. The two-dimensional axisymmetric simulation domain and boundaries setup are shown in the inset of \cref{fig:validation}a. The salt concentration of the rightmost reservoir boundary was determined by \cref{eq:Soret}. The induced potential $ \Delta \varphi $ was determined by the constraint of $ I=0 $. Such constraint was carried out by using a \texttt{Global ODE} module in our simulation. Additionally, at the capillary wall a constant surface charge density ($ \bar{q} $) condition was employed. To relate $ \bar{q} $ (normalized by $ \epsilon_{\mathrm{ref}} k_{\mathrm{B}}T_{\mathrm{ref}}/(ea)$) to $ \bar{\zeta} $, for given $ \bar{\zeta} $ and $ \bar{\kappa}_0 $ \cref{eq:pbe} was numerically solved with \cref{eq:pb_bc} for a set of successive values $ \theta_i \in [0,1] $. Then a set of data $ \bar{q}_i $ corresponding to $ \theta_i $ was obtained, and was further fitted by a quadratic function as $ \bar{q}_{\bar{\zeta},\bar{\kappa}_0}=\bar{q}_{\bar{\zeta},\bar{\kappa}_0}(\theta) $ (\cref{fig:validation}a, inset). In the capillary, a structured rectangular mesh with 2000 elements in $ x $ direction and 100 elements in $ r $ direction was employed, whereas in the reservoirs an extremely fine triangular mesh was employed. In addition, an element ratio of at least 10 was adopted to refine the mesh density at capillary wall. The independence of current computing grid has been well verified.

\section{Derivation of the maximum efficiency}\label{appD}
For non-vanishing average current density, \cref{eq:varphi_diff}~can be further extended as
\begin{widetext}
\begin{align}\label{eq:iv}
\frac{e\Delta\varphi}{k_{\mathrm{B}} T_{\mathrm{ref}}}=-\frac{eS_e \Delta T}{k_{\mathrm{B}}T_{\mathrm{ref}}} -\frac{I}{en_{\infty}u_{\mathrm{ref}}\Delta T}  \int_{T_{\infty}}^{T_{\infty}+\Delta T} \left( \frac{u_{\mathrm{ref}}l}{4D} \frac{T}{T_{\mathrm{ref}}}\frac{1}{\beta L_6+L_7}+\frac{A_0 B_0}{4} \frac{\mu}{\mu_{\mathrm{ref}}} \mathscr{A} \right) \frac{n_{\infty}}{n}  \mathrm{d}T 
\end{align}	
\end{widetext}

with
\begin{equation}\label{eq:curr_coeff}
\mathscr{A}=\frac{\int_{T_{\infty}}^{T_{\infty}+\Delta T} \frac{A_0 B_0}{4} \frac{\mu}{\mu_{\mathrm{ref}}} \frac{n_{\infty}}{n} \mathrm{d}T }{\int_{T_{\infty}}^{T_{\infty}+\Delta T} \frac{\mu}{\mu_{\mathrm{ref}}} B_0 \mathrm{d}T } -\frac{A_0}{4}\frac{n_{\infty}}{n}
\end{equation}
Setting $ \Delta\varphi=0 $, one obtains the average short-circuit current density as
\begin{equation}\label{eq:sc_current}
I_{|\Delta \varphi=0}\approx -\frac{2 S_e \Delta T}{\int_{T_{\infty}}^{T_{\infty}+\Delta T} \frac{k_{\mathrm{B}}T}{2e^2Dn} \frac{\mathrm{d}T}{\beta L_6+L_7}}\frac{\Delta T}{l}
\end{equation}
by noting that $ \int_{T_{\infty}}^{T_{\infty}+\Delta T} (A_0 B_0/4) (\mu/\mu_{\mathrm{ref}})(n_{\infty}/n) \mathscr{A} \mathrm{d}T \approx 0$.

Due to the linear current-voltage relationship (\ref{eq:iv}), the maximum output power density is derived as $ P_{\mathrm{o}}=I_{|\Delta T=0} \Delta\varphi_{|I=0}/4 $. The heat power density is approximated by $ P_{\mathrm{i}} \approx k_{\mathrm{ave}}\Delta T/l $. Then, the maximum efficiency can be derived as
\begin{equation}\label{eq:efficiency}
\eta_e=\frac{P_{\mathrm{o}}}{P_{\mathrm{i}}}\approx \frac{ (S_e \Delta T)^2}{2 k_{\mathrm{ave}}\int_{T_{\infty}}^{T_{\infty}+\Delta T}  \frac{\sigma_{\infty}^{-1}\mathrm{d}T}{\beta L_6+L_7}}
\end{equation}
where $ \sigma_{\infty} = 2e^2 n D  / (k_{\mathrm{B}}T)  $ is the local bulk electrical conductivity of the electrolyte solution. Since $ \int_{T_{\infty}}^{T_{\infty}+\Delta T}\sigma_{\infty}^{-1} / (\beta L_6+L_7)\mathrm{d}T \approx  [\sigma_{\infty}^{-1}/(\beta L_6+L_7)]_{\mathrm{ave}}\Delta T $, \cref{eq:efficiency} can be further simplified as
\begin{equation}\label{eq:reduced_efficiency}
\eta_e\approx ZT_{\mathrm{ave}} \frac{T_{\infty}+\Delta T}{2T_{\mathrm{ave}}}\eta_{\mathrm{c}}
\end{equation}
where $ ZT_{\mathrm{ave}}=[(\beta L_6+L_7)\sigma_{\infty}S_e^2 T/k]_{\mathrm{ave}} $ is the average figure of merit and $ \eta_{\mathrm{c}}=\Delta T/(T_{\infty}+\Delta T) $ is the Carnot efficiency.

%
\bibliography{hmt19.bib}

\end{document}